\documentstyle[epsfig,12pt]{article}

\begin{document}
\begin{center}
{\Large\bf Covariant tensor formalism for partial wave analyses of
$\psi$ decays into $\gamma B\bar B$, $\gamma\gamma V$ and
$\psi(2S)\to\gamma\chi_{c0,1,2}$ with $\chi_{c0,1,2}\to  K\bar K
\pi^+\pi^- $ and
 $2\pi^+2\pi^-$} \vskip 0.5cm

{\large\bf Sayipjamal
Dulat$^{a,b)}$, Bing-Song Zou$^{c,a)}$}\\
\vskip 0.5cm a) Institute of High Energy Physics, CAS, P.O.~Box
918(4), Beijing 100049, China\\
b) Department of Physics, Xinjiang University, Urumqi 830046   China \\
c) CCAST (World Laboratory), P.O.~Box 8730, Beijing 100080, China\\

\vskip 0.5cm
\end{center}

\begin{abstract}
With accumulation of high statistics data at BES and CLEO-c, many
new interesting channels can get enough statistics for partial
wave analysis (PWA). Among them,  $\psi \to \gamma p\bar p,
\gamma\Lambda\bar \Lambda, \gamma\Sigma \bar\Sigma, \gamma\Xi\bar
\Xi$ channels provide a good place for studying baryon-antibaryon
interactions; the double radiative decays $\psi\to\gamma\gamma V$
with $V\equiv \rho,\omega,\phi$ have a potential to provide
information on the flavor content of any meson resonances (R) with
positive charge parity ($C=+$) and mass above 1 GeV through
$\psi\to\gamma R \to\gamma\gamma V$;
$\psi(2S)\to\gamma\chi_{c0,1,2}$ with $\chi_{c0,1,2}\to  K\bar K
\pi^+\pi^- $ and $2\pi^+2\pi^-$ decays are good processes to study
$\chi_{cJ}$ charmonium decays. Using the covariant tensor
formalism, here we provide theoretical PWA formulae for these
channels.
\end{abstract}

\section{Introduction}
Abundant $J/\psi$ and $\psi'$ events have been collected at the
Beijing Electron Positron Collider (BEPC). More data will be
collected at upgraded BEPC and CLEO-C. Many new interesting
channels are now getting enough statistics for partial wave
analysis.

$J/\psi$ and $\psi'$ radiative decay to $B\bar B$ (baryon and
antibaryon pair) is a good place to study baryon-antibaryon
interactions and to look for possible resonant states of the
$B\bar B$ system. Based on the 58 million $J/\psi$ events
accumulated by the BES2 detector at the BEPC, recently BES2
reported \cite{BES} that they observed a strong, narrow
enhancement near the threshold in the invariant mass spectrum of
$p\bar p$ (proton - antiproton) pairs from $J/\psi\to \gamma p\bar
p$ radiative decays.  The structure has attracted people's
attention to study $p\bar p$ near-threshold interaction
\cite{Chiang}. Future data on $\psi \to\gamma\Lambda\bar \Lambda,
\gamma\Sigma \bar\Sigma, \gamma\Xi\bar \Xi$ channels will give a
new opportunity to study hyperon-antihyperon interactions.

$J/\psi$ and $\psi'$ double radiative decays $\psi\to\gamma\gamma
V(\rho, \omega, \phi)$ provide a favorable place to extract the
$u\bar u$, $d\bar d$ and $s\bar s$ structure of intermediate
states \cite{Close}. The $J/\psi\to\gamma\gamma\rho$ and
$\gamma\gamma\phi$ have been studied  by Crystal Ball \cite{ed},
DM2 \cite{aug}, MARK-III \cite{coff} and BES-I \cite{BES0}. An
interesting structure at $\iota(1440)$ region in the $\gamma V$
invariant mass spectra is observed. But due to limited statistics,
one cannot get reliable PWA results. With much higher statistics
$\psi$ data to be available soon at CLEO-C and BES-III, these
$\psi$ double radiative decay channels give a potential to provide
information on the flavor content of any meson resonances (R) with
positive charge parity ($C=+$) and mass above 1 GeV through
$\psi\to\gamma R \to\gamma\gamma V$.

The $\psi(2s)$ radiative decays into $K^+K^-\pi^+\pi^-$ and
$\pi^+\pi^-\pi^+\pi^-$ via $\chi_{cJ}$ intermediate states are
good processes to study $\chi_{cJ}$ decays which may provide
useful information on two-gluon hadronization dynamics and
glueball decays.

In order to get more useful information about
 the resonance properties such as their $J^{PC}$ quantum
numbers, mass, width, production and decay rates, etc., partial
wave analyses (PWA) are necessary. PWA is an effective method for
analysing the experimental data of hadron spectrum. There are two
types of PWA: one is based on the covariant tensor (also named
Rarita-Schwinger) formalism\cite{Rarita} and the other is based on
the helicity formalism\cite{Chung}. Ref.\cite{filip} showed the
connection between the covariant tensor formalism and the
 helicity one. Ref.\cite{zou} provided PWA formulae in the
covariant tensor formalism for $\psi$ decays to mesons, which have
been used for a number of channels already published by
BES\cite{BES-PWA} and are going to be used for more channels. A
similar approach has been used in analyzing other
reactions\cite{Anisovich1,Anisovich2,Anisovich3}. Ref.\cite{zou2}
provided explicit formulae for the angular distribution of the
photon in $\psi$ radiative decays in the covariant tensor
formalism, and also discussed  helicity formalism of the angular
distribution of the $\psi$ radiative decays to two pseudoscalar
mesons, and its relation to the covariant tensor formalism.

In this paper we extend the covariant tensor formalism \cite{zou}
to give explicit PWA formulae for the new interesting channels
mentioned above. The plan of this article is as follows: in
section 2, we present the necessary tools for the calculation of
the tensor amplitudes, within a covariant tensor formalism. This
will allow us to derive covariant amplitudes for all possible
processes.  In section 3, we present covariant tensor formalism
for $\psi$ radiative decays to baryon antibaryon pairs. In section
4, we present covariant tensor formalism for $\psi$ decays into
$\gamma\gamma V(\rho,\omega,\phi)$. In section 5, we present
covariant tensor formalism for the $\psi(2s)$ decays into $\gamma
K^+ K^- \pi^+\pi^-$ and $\gamma \pi^+\pi^-\pi^+\pi^-$,
respectively. The conclusions are given in section 6. Since
covariance is a useful property of any decay amplitude, all
possible amplitudes are written in terms of covariant tensor form.
All amplitudes include a complex coupling constant and
Blatt-Weisskopf centrifugal barriers where necessary.

\section{Prescriptions for the construction of covariant tensor
amplitudes}

 In this section we present the necessary tools for the
construction of covariant tensor amplitudes. Following the
convention of Ref.\cite{zou} for the $\psi$ decays, the partial
wave amplitudes $U_i^{\mu\nu\alpha}$ in the covariant
Rarita-Schwinger tensor formalism  can be constructed by using
pure orbital angular momentum covariant tensors $\tilde
t^{(L_{bc})}_{\mu_1\cdots\mu_{L_{bc}}}$ and covariant spin wave
functions $\phi_{\mu_1 \cdots\mu_s}$ together with the  metric
tensor $g^{\mu\nu}$,  the totally antisymmetric Levi-Civita tensor
$\epsilon_{\mu\nu\lambda\sigma}$ and the four momenta of
participating particles; here the indices $\mu$,$\nu$, and
$\alpha$ run from $1$ to $4$ over $x$, $y$, $z$, and $t$.
 For a process $a\to bc$, if there exists a relative orbital
angular momentum $\textbf{L}_{bc}$ between the particle b and c,
then the
 pure orbital angular momentum
 $\textbf{L}_{bc}$ state can be represented by the
covariant tensor wave function $\tilde
t^{(L_{bc})}_{\mu_1\cdots\mu_{L_{bc}}}$ \cite{Chung} which is
built out of the relative momentum. Thus here we give only
covariant tensors that correspond to the pure S-, P-, D-, and
F-wave orbital angular
 momenta:
\begin{eqnarray}\label{l0}
\tilde t^{(0)} &=& 1, \\\label{l1}
 \tilde t^{(1)}_\mu &=& \tilde
g_{\mu\nu}(p_a)r^\nu B_1(Q_{abc}) \equiv\tilde r_\mu
B_1(Q_{abc}),\\\label{l2} \tilde t^{(2)}_{\mu\nu} &=& [\tilde
r_\mu\tilde r_\nu -{1\over 3}(\tilde r\cdot\tilde r)\tilde
g_{\mu\nu}(p_a)]B_2(Q_{abc}), \\\label{l3} \tilde
t^{(3)}_{\mu\nu\lambda} &=& [\tilde r_\mu\tilde r_\nu\tilde
r_\lambda -{1\over 5}(\tilde r\cdot\tilde r)(\tilde
g_{\mu\nu}(p_a)\tilde r_\lambda +\tilde g_{\nu\lambda}(p_a)\tilde
r_\mu+\tilde g_{\lambda\mu}(p_a)\tilde r_\nu)]B_3(Q_{abc}),
\\ & &
\cdots  \nonumber \\
p_a^\mu \tilde t^{(1)}_\mu &= &p_a^\mu \tilde t^{(2)}_{\mu\nu}
=p_a^\mu t^{(3)}_{\mu\nu\lambda} =  0
\end{eqnarray}
where $r=p_b-p_c$  is the relative  four momentum of the two decay
products in the parent particle rest frame; $(\tilde r\cdot\tilde
r)= -{\textbf{r}^2}$. and
\begin{equation}
\tilde g_{\mu\nu}(p_a) = g_{\mu\nu} -
\frac{p_{a\mu}p_{a\nu}}{p_a^2};
\end{equation}
Here the Minkowsky metric tensor has the form
$$
g_{\mu\nu} = diag(1,-1,-1,-1).
$$
$B_{L_{bc}}(Q_{abc})$ is a Blatt-Weisskopf barrier
factor\cite{Chung,Hippel}, where $Q_{abc}$ is the magnitude of
$\bf p_b$ or $\bf p_c$ in the rest system of $a$,
\begin{equation}
Q_{abc}^2=\frac{(s_{a}+s_{b}-s_{c})^{2}}{4s_{a}}-s_{b}
\end{equation}
with $s_a=E_a^2-{\bf p}^2_a$.

The spin-1 and spin-2 particle wave functions $\phi_\mu(p_a,m_s)$
and $\phi_{\mu\nu}(p_a,m_s)$ with spin projection  $m_s$ satisfy
the following conditions
\begin{eqnarray}
&& p_a^{\mu} \phi_\mu(p_a,m_s)=0,
\hspace{1cm}\phi_\mu(p_a,m_s)\phi^{*\mu}(p_a,m'_s)=-\delta_{m_s
 m'_s}\nonumber\\\label{proj1} &&
\sum_m\phi_\mu(p_a,m_s)\phi^{*}_{\nu}(p_a,m_s)=-g_{\mu\nu}+\frac{p_{a\mu}p_{a\nu}}{
p_a^2}\equiv -\tilde g_{\mu\nu}(p_a),\\\nonumber
&&p_a^\mu\phi_{\mu\nu}(p_a,m_s)=0,\hspace{0.8cm}\phi_{\mu\nu}=\phi_{\nu\mu},\hspace{0.8cm}
 g^{\mu\nu}\phi_{\mu\nu}=0,\hspace{0.8cm}\\
&&\phi_{\mu\nu}(p_a,m_s)\phi^{*\mu\nu}(p_a,m'_s)=\delta_{m_s
m'_s}\nonumber\\\label{proj2}\nonumber
\end{eqnarray}
Projection operators will be a useful general tool in constructing
expressions. The spin-2 projection operator has the
form\cite{Chung,zou}
\begin{equation}
P^{(2)}_{\mu\nu\mu'\nu'}(p_a) =
\sum_m\phi_{\mu\nu}(p_a,m_s)\phi^*_{\mu'\nu'}(p_a,m_s)={1\over 2}
(\tilde g_{\mu\mu'}\tilde g_{\nu\nu'}+\tilde g_{\mu\nu'}\tilde
g_{\nu\mu'})-{1\over 3}\tilde g_{\mu\nu}\tilde g_{\mu'\nu'}.
\end{equation}

Note that for a given decay process $a\to bc$, the total angular
momentum should be conserved, which means
\begin{equation}\label{conservation}
\textbf{J}_a = \textbf{S}_{bc} + \textbf{L}_{bc},
\end{equation}
where
\begin{equation}
\textbf{S}_{bc} = \textbf{S}_b + \textbf{S}_c .
\end{equation}
In addition parity should also be conserved, which means
\begin{equation}
\eta_a = \eta_b\eta_c (-1)^{L_{bc}},
\end{equation}
where $\eta_a$,  $\eta_b$, and $\eta_c$ are the intrinsic parities
of particles $a$, $b$, and $c$, respectively. From this relation,
one knows whether $L_{bc}$ should be even or odd. Then from Eq.
(\ref{conservation}) one can find out how many different
$(L_{bc},S_{bc})$ combinations there are, which determine the
number of independent couplings. Also note that in the
construction of the covariant tensor amplitude,  if
$S_{bc}+L_{bc}+J_a$ is an odd number, then
$\epsilon_{\mu\nu\lambda\sigma}p_a^\sigma$ with $p_a$ the momentum
of the parent particle is needed; otherwise it is not needed. See,
for example, Eq. (\ref{1p1}) below.

\section{Covariant tensor formalism for $\psi$ decay  into
 $\gamma B\bar B$}
The general form of the decay $\psi \rightarrow \gamma X
\rightarrow \gamma p\bar p$ amplitude can be written as follows by
using the polarization four-vectors of the initial and final
states,
\begin{eqnarray}
A^{(s)}&=&\psi_\mu(p,m_J)
e^*_\nu(q,m_\gamma)\psi_{\alpha_s}(p_b,S_b; p_c,S_c)A^{\mu\nu\alpha_s}\nonumber\\
 &=&\psi_\mu(p,m_J)
e^*_\nu(q,m_\gamma)\psi_{\alpha_s}(p_b,S_b;p_c,S_c)\sum_i\Lambda_i
U_i^{\mu\nu\alpha_s}.
\end{eqnarray}
where $\psi_\mu(p,m_J)$ is the polarization four vector of the
$\psi$ with spin projection of $m_J$; $e_\nu(q,m_\gamma)$ is the
polarization four vector of the photon with spin projections of
$m_\gamma$; $\psi_{\alpha_s}(p_b,S_b; p_c,S_c)$ is the spin wave
function of the proton and antiproton system with spin $S_b$ and
$S_c$, respectively, and the index $s$ stands for the total spin
 of the $p \bar p$, see, for example, Eq. (\ref{pps0},\ref{pps1});
 $U_i^{\mu\nu\alpha_s}$ is the $i$-th partial wave amplitude with
coupling strength determined by a complex parameter $\Lambda_i$.
The spin-1 polarization vector $\psi_\mu(p,m_J)$ for $\psi$ with
four momentum $p_\mu$ satisfies
\begin{equation}\label{spin1}
\sum_{m_{J}=1}^3\psi_\mu(p,m_J)\psi^{*}_{\nu}(p,m_J)=
-g_{\mu\nu}+{p_\mu p_\nu\over p^2}\equiv -\tilde g_{\mu\nu}(p),
\end{equation}
with $p^\mu\psi_\mu  = 0$. For $\psi$ production from $e^+ e^-$
annihilation, the electrons are highly relativistic, with the
result that $J_z = \pm 1$ for the $\psi$ spin projection taking
the beam direction as the z-axis. This limits $m_J$ to 1 and 2,
i.e. components along $x$ and $y$. Then one has the following
relation
\begin{equation}\label{jpsi}
\sum^2_{m_J=1}\psi_{\mu}(p,m_J)\psi^*_{\mu'}(p,m_J)
=\delta_{\mu\mu'}(\delta_{\mu 1}+\delta_{\mu 2}).
\end{equation}
For the  photon polarization four vector, there is the usual
Lorentz orthogonality condition. Namely, the polarization four
vector $e_\nu(q,m_\gamma)$ of the  photon with  momenta $q$
satisfies
\begin{equation}\label{photono}
q^\nu e_\nu(q,m_\gamma)=0,
\end{equation}
which states that spin-1 wave function is orthogonal to its own
momentum.
 The above relation is the same as for a massive vector meson. However, for
the photon, there is an additional gauge invariance condition.
Here we assume the Coulomb gauge in the $\psi$ rest system, {\sl
i.e.}, $p^\nu e_\nu =0$.  Then we have \cite{Greiner}
\begin{equation}\label{photon}
\sum_{m_\gamma} e^*_\mu(q,m_\gamma) e_\nu(q,m_\gamma) =
-g_{\mu\nu}+\frac{q_\mu K_\nu+ K_\mu q_\nu}{q\!\cdot\!
K}-\frac{K\!\cdot\! K}{(q\!\cdot\! K)^2}q_\mu q_\nu \equiv
-g^{(\perp\perp)}_{\mu\nu}
\end{equation}
with $K=p-q$ and $K^\nu e_\nu =0$. We denote the four momentum of
the particle $X$ by $K$, and $q\!\cdot\! K$ is a four vector dot
product.
 For $X \to p\bar p $,
the total spin of $p\bar p$ system can be either $0$ or $1$. These
two states can be represented by $\psi$ and $\psi_\alpha$
\cite{zou1}. where
\begin{eqnarray}\label{pps0}
 \psi &= & \bar u(p_b,S_b)\gamma_5
v(p_c,S_c), \hspace{4cm} if \hspace{0.3cm} s=0 ,\\ \label{pps1}
\psi_{\alpha} &=&\bar u(p_b,S_b)(\gamma_\alpha -
\frac{r_\alpha}{m_X + 2m} )v(p_c,S_c), \hspace{1.2cm} if
\hspace{0.5cm} s=1 .
\end{eqnarray}
One can see that both  $\psi$ and $\psi_\alpha$ have no dependence
on the direction of the momentum $\hat{\textbf{p}}$, hence
correspond to pure spin states with the total spin of $0$ and $1$,
respectively. Here $p_b$, $p_c$, and $S_b$, $S_c$ are momenta and
spin of the proton antiproton pairs, respectively. $m_X$ and $m$
are the masses of $X$ and  $p$, $\bar p$, respectively;
$u(p_b,S_b)$ and $v(p_c,S_c)$ are the standard Dirac spinors. If
we sum over the polarization, we have the two projection
operators:
\begin{eqnarray}\label{diracpo}
\sum_{S_b}u_\alpha(p_b,S_b)\bar u_\beta(p_b,S_b)&=&
\Big(\;\frac{\not\!p_b +
m }{2m}\Big)_{\alpha\beta}\nonumber\\
\sum_{S_c}v_\alpha(p_c,S_c)\bar v_\beta(p_c,S_c)&=&
\Big(\;\frac{\not\!p_c - m }{2m}\Big)_{\alpha\beta}
\end{eqnarray}

To compute the differential cross section, we need an expression
for $|A|^2$. Note that the square modulus of the decay amplitude,
which gives the decay probability  of a certain configuration
should be independent of any particular frame, that is, a Lorentz
scalar. Thus by using  Eqs. (\ref{jpsi}) and (\ref{photon}), the
differential cross section for the radiative decay to 3-body final
state is:
\begin{eqnarray}
\frac{d\sigma^{(s)}}{d\Phi_3}\!&=&\!\frac{1}{2}\sum_{S_b,S_c}\sum^2_{m_J=1}\sum^2_{m_\gamma=1}
|\psi_{\mu}(p,m_J)
e^*_{\nu}(q,m_\gamma)\psi_{\alpha_s}(p_b,S_b;p_c,S_c)A^{\mu\nu\alpha_s}|^2
\nonumber\\
&=& -\frac{1}{2}\sum_{S_b,S_c}\sum^2_{\mu=1}A^{\mu\nu\alpha_s}
g^{(\perp\perp)}_{\nu\nu'}A^{*\mu\nu'\alpha'_s}
\psi^{*}_{\alpha_s}\psi_{\alpha'_s}\nonumber\\
&=& -\frac{1}{2}\sum_{i,j}\Lambda_i\Lambda_j^*\sum^2_{\mu=1}
U_i^{\mu\nu\alpha_s}g^{(\perp\perp)}_{\nu\nu'}U_j^{*\mu\nu'\alpha'_s}
\sum_{S_b,S_c}\psi^{*}_{\alpha_s}\psi_{\alpha'_s}\nonumber\\
 &\equiv& \sum_{i,j}P_{ij}\cdot F^{(s)}_{ij}
\end{eqnarray}
where
\begin{eqnarray}
P_{ij} &= P^*_{ji} &= \Lambda_i\Lambda^*_j, \nonumber\\
F^{(s)}_{ij} &= F^{*(s)}_{ji} &= -\frac{1}{2}\sum^2_{\mu=1}
U_i^{\mu\nu\alpha_s}g^{(\perp\perp)}_{\nu\nu'}U_j^{*\mu\nu'\alpha'_s}
\sum_{S_b,S_c}\psi^{*}_{\alpha_s}\psi_{\alpha'_s}.
\end{eqnarray}
$d\Phi_3$ is the standard Lorentz invariant 3-body phase space
given by
\begin{equation}
d\Phi_3(p; q, p_b, p_c)=\delta^4(p - q - p_b - p_c) \frac{d^3{\bf
q}}{(2\pi)^3 2E_\gamma }  \frac{m^2 d^3{\bf p}_b d^3{\bf p}_c }
{(2\pi)^3 E_b (2\pi)^3 E_c }.
\end{equation}
\begin{eqnarray}
F^{(0)}_{ij} &=& F^{*(0)}_{ji} = -\frac{1}{2}\sum^2_{\mu=1}
U_i^{\mu\nu}g^{(\perp\perp)}_{\nu\nu'}U_j^{*\mu\nu'}
\sum_{S_b,S_c}\psi^{*}\psi\nonumber\\
&=&\frac{1}{2}\sum^2_{\mu=1}
U_i^{\mu\nu}g^{(\perp\perp)}_{\nu\nu'}U_j^{*\mu\nu'}
Tr\Big(\;\frac{\not\!p_b + m }{2m}\;\gamma_5 \;\frac{\not\!p_c - m
}{2m}\;\gamma_5 \;\Big)\nonumber\\
&=&-\frac{m^{2}_X}{4m^2}\sum^2_{\mu=1}
U_i^{\mu\nu}g^{(\perp\perp)}_{\nu\nu'}U_j^{*\mu\nu'}.
\end{eqnarray}
The spin sums can be performed using the completeness relations
from Eq.  (\ref{diracpo}):
\begin{eqnarray}
F^{(1)}_{ij} &= &F^{*(1)}_{ji} = -\frac{1}{2}\sum^2_{\mu=1}
U_i^{\mu\nu\alpha}g^{(\perp\perp)}_{\nu\nu'}U_j^{*\mu\nu'\alpha'}
\sum_{S_b,S_c}\psi^{*}_\alpha\psi_{\alpha'}\nonumber\\
&=& -\frac{1}{2} \sum^2_{\mu=1}
U_i^{\mu\nu\alpha}g^{(\perp\perp)}_{\nu\nu'}U_j^{*\mu\nu'\alpha'}
 \Big[ Tr\Big(\;\frac{\not\!p_b + m }{2m}\;\gamma_\alpha
\;\frac{\not\!p_c - m }{2m}\;\gamma_{\alpha'}\;\Big)\nonumber\\
&-& \frac{r_\alpha}{m_X + 2m} Tr\Big(\;\frac{\not\!p_b + m
}{2m}\;\frac{\not\!p_c - m }{2m}\;\gamma_{\alpha'}\;\Big) -
\frac{r_{\alpha'}}{m_X + 2m} Tr\Big(\;\frac{\not\!p_b + m
}{2m}\;\gamma_\alpha \;\frac{\not\!p_c - m }{2m}\;\Big)\nonumber\\
& + & \frac{r_\alpha r_{\alpha'}}{(m_X  + 2m)^2}
Tr\Big(\;\frac{\not\!p_b + m }{2m}\;\frac{\not\!p_c - m
}{2m}\;\Big) \Big]\nonumber\\
&=& -\frac{1}{4m^2} \sum^2_{\mu=1}
U_i^{\mu\nu\alpha}g^{(\perp\perp)}_{\nu\nu'}U_j^{*\mu\nu'\alpha'}
 \Big[p_{b\alpha} p_{b\alpha'}+p_{c\alpha} p_{c\alpha'}+
 p_{b\alpha} p_{c\alpha'} +  p_{b\alpha'} p_{c\alpha}
 - m^{2}_X g_{\alpha\alpha'}
 \Big].\nonumber\\
\end{eqnarray}

\subsection*{Amplitudes for the radiative decay
$\psi  \rightarrow \gamma p\bar p$}

We consider the decay of a $\psi$ state in two steps:
$\psi\to\gamma X$ with $X\to p\bar p$. The possible $J^{PC}$ for
$X$ are $0^{++}, 0^{-+}$, $1^{++}$, $2^{++}$, $2^{-+}$, etc. For
$\psi\to\gamma X$, we choose two independent momenta $p$ for
$\psi$ and $q$ for the photon to be contracted with spin wave
functions. We denote the four momentum of $X$  by $K$. The tensor
describing the first and second steps will be denoted by $\tilde
T^{(L)}_{\mu_1\cdots\mu_L}$ and $\tilde
t^{(\textit{l})}_{\mu_1\cdots\mu_{\textit{l}}}$, respectively.

For $\psi\to\gamma 0^{++} \rightarrow \gamma p\bar p$, there is
one independent covariant tensor amplitude:
\begin{equation}\label{0p}
U^{\mu\nu\alpha} = g^{\mu\nu}  \tilde t^{(1)\alpha}.
\end{equation}

For $\psi\to\gamma 0^{-+} \rightarrow \gamma p\bar p$, there is
one independent covariant tensor amplitude:
\begin{equation}\label{0m}
U^{\mu\nu}=  \epsilon^{\mu\nu\lambda\sigma}p_\lambda q_\sigma
B_1(Q_{\psi\gamma X}).
\end{equation}

 For $\psi\to\gamma 1^{++} \rightarrow
\gamma p\bar p$, there are two independent covariant tensor
amplitudes:
\begin{eqnarray}\label{1p1}
U^{\mu\nu\alpha}_1 \!&=&\! \epsilon^{\mu\nu\lambda\sigma}p_\lambda
\epsilon^{\alpha\beta\rho}_{\;\;\;\;\;\;\sigma}K_\rho \tilde
t^{(1)}_\beta  ,\\\label{1p2}U^{\mu\nu\alpha}_ 2 \!&=&\!
\epsilon^{\nu\lambda\sigma\gamma}p_\lambda q^{\mu} q_{\gamma}
\epsilon^{\alpha\beta\rho}_{\;\;\;\;\;\;\sigma}K_\rho \tilde
t^{(1)}_\beta B_{2}(Q_{\psi\gamma X}).
\end{eqnarray}

For $\psi\to\gamma 1^{-+}$, the exotic $1^{-+}$ meson cannot decay
into $p\bar p$.

For $\psi\to\gamma 2^{++} \rightarrow  \gamma p\bar p$, there are
six independent covariant tensor amplitudes:
\begin{eqnarray}\label{2p1}
U^{\mu\nu\alpha}_1\!&=&\! P^{(2)\mu\nu\alpha\beta}(K)\tilde
t^{(1)}_\beta ,\\\label{2p2} U^{\mu\nu\alpha}_2  \!&=&\!
P^{(2)\mu\nu\lambda\beta}(K) \tilde t^{(3)\alpha}_{\lambda\beta
},\\\label{2p3} U^{\mu\nu\alpha}_3\!&=&\!
P^{(2)\nu\sigma\alpha\beta}
 q^{\mu}p_{\sigma}
\tilde t^{(1)}_\beta B_{2}(Q_{\psi\gamma X}),\\\label{2p4}
U^{\mu\nu\alpha}_4\!&=&\! P^{(2)\nu\sigma\lambda\beta}
 q^{\mu}p_{\sigma}
\tilde t^{(3)\alpha}_{\lambda\beta} B_{2}(Q_{\psi\gamma
X}),\\\label{2p4} U^{\mu\nu\alpha}_5\!&=&\!
 g^{\mu\nu} P^{(2)\sigma\rho\alpha\beta}
p_{\sigma}p_{\rho}
 \tilde t^{(1)}_\beta B_{2}(Q_{\psi\gamma X})
,\\\label{2p5} U^{\mu\nu\alpha}_6\!&=&\!
 g^{\mu\nu} P^{(2)\sigma\rho\lambda\beta}
p_{\sigma}p_{\rho}
 \tilde t^{(3)\alpha}_{\lambda\beta }B_{2}(Q_{\psi\gamma X})
\end{eqnarray}
where $\tilde t^{(1)}$ and $\tilde t^{(3)}$ correspond to the
orbital angular momentum between the proton and antiproton $l$ to
be 1 and 3, respectively.

For $\psi\to\gamma 2^{-+} \rightarrow  \gamma p\bar p$,  the
possible partial wave amplitudes are the following:
\begin{eqnarray}\label{2m1}
U^{\mu\nu}_ 1 \!&=&\! \epsilon^{\mu\nu\lambda\sigma}p_\lambda
q^{\gamma} \tilde t^{(2)}_{ \gamma\sigma} B_{1}(Q_{\psi\gamma X}),
 \\\label{2m2}
U^{\mu\nu}_ 2 \!&=&\! \epsilon^{\mu\nu\lambda\sigma}p_\lambda
q_{\sigma}p_{\gamma}p_{\delta} \tilde t^{(2)\gamma\delta}
B_{3}(Q_{\psi\gamma X}),
 \\\label{2m3}
U^{\mu\nu}_ 3 \!&=&\! \epsilon^{\nu\gamma\lambda\sigma}p_\lambda
q_{\sigma}q^{\mu}p^{\delta} \tilde t^{(2)}_{\gamma\delta}
B_{3}(Q_{\psi\gamma X}).
\end{eqnarray}

It is worth  mentioning here that the above partial wave
amplitudes for the process $J/\psi \to \gamma p\bar p$ are
applicable to the processes $J/\psi \to \gamma\Lambda\bar \Lambda,
\gamma\Sigma \bar\Sigma$, and $\gamma\Xi\bar \Xi$ as well.

\section{Covariant tensor formalism for $\psi$ decay  into
 $\gamma \gamma V$}

By using the polarization four-vectors of the initial and final
states, now  we write the general form of the decay amplitude for
the process
\begin{equation}\label{psivv}
\psi \rightarrow \gamma R \rightarrow \gamma \gamma V(\rho, \phi,
\omega)
\end{equation}
as follows
\begin{equation}
A=\psi_\mu(p,m_J) e^*_\nu(q,m_\gamma)
\varepsilon^*_\alpha(k,m'_\gamma) A^{\mu\nu\alpha}
=\psi_\mu(p,m_J)
e^*_\nu(q,m_\gamma)\varepsilon^*_\alpha(k,m'_\gamma)\sum_i\Lambda_i
U_i^{\mu\nu\alpha}.
\end{equation}
In the following $e_\nu(q,m_\gamma)$ denotes the polarization
function of the photon in $\psi \rightarrow \gamma R$, and
$\varepsilon_\alpha(k,m'_\gamma)$ denotes that of the photon in $R
\to \gamma V$. The polarization four vectors $\psi_\mu(p,m_J)$ and
$e_\nu(q,m_\gamma)$  satisfy the conditions
$(\ref{spin1}-\ref{photon})$.  And
$\varepsilon_\alpha(k,m'_\gamma)$  satisfy
\begin{equation}
k^\alpha\varepsilon_\alpha(k, m'_{\gamma})=0,
\end{equation}
\begin{equation}\label{photon2}
\sum_{m'_{\gamma}} \varepsilon^*_\alpha(k,m'_{\gamma})
\varepsilon_\beta(k,m'_{\gamma}) = -g_{\alpha\beta}+\frac{k_\alpha
p_{V\beta}+ p_{V\alpha} k_\beta}{k\!\cdot\!
p_V}-\frac{p_V\!\cdot\! p_V}{(k\!\cdot\! p_V)^2}k_\alpha k_\beta
\equiv -g^{(\perp)}_{\alpha\beta}
\end{equation}
with $p_V=K-k$ and $p_V^\alpha\varepsilon_\alpha =0$. We denote
the four momenta of the particles $R$ and $V(\rho, \phi, \omega)$
by $K$ and $p_V$, respectively. Then the differential cross
section for the radiative decay to an n-body final state is:
\begin{eqnarray}
\frac{d\sigma}{d\Phi_n}\!&=&\!\frac{1}{2}\sum^2_{m_J=1}\sum^3_{m'_\gamma,m_\gamma=1}
|\psi_{\mu}(p,m_J)
e^*_{\nu}(q,m_\gamma)\varepsilon^*_{\alpha}(k,m'_\gamma)A^{\mu\nu\alpha}|^2
\nonumber\\ \!&=&\!\frac{1}{2}\sum^2_{m_J=1}
\psi_{\mu}(p,m_J)\psi^*_{\mu'}(p,m_J)
g^{(\perp\perp)}_{\nu\nu'}g^{(\perp)}_{\alpha\alpha'}
A^{\mu\nu\alpha}A^{*\mu'\nu'\alpha'}\nonumber\\
&=& \frac{1}{2}\sum^2_{\mu=1}A^{\mu\nu\alpha}
g^{(\perp\perp)}_{\nu\nu'}g^{(\perp)}_{\alpha\alpha'}A^{*\mu\nu'\alpha'}\nonumber\\
&=& \frac{1}{2}\sum_{i,j}\Lambda_i\Lambda_j^*\sum^2_{\mu=1}
U_i^{\mu\nu\alpha}g^{(\perp\perp)}_{\nu\nu'}g^{(\perp)}_{\alpha\alpha'}
U_j^{*\mu\nu'\alpha'}
 \equiv \sum_{i,j}P_{ij}\cdot F_{ij}
\end{eqnarray}
where
\begin{eqnarray}
P_{ij} &= P^*_{ji} &= \Lambda_i\Lambda^*_j, \\
F_{ij} &= F^*_{ji} &= \frac{1}{2}\sum^2_{\mu=1}
U_i^{\mu\nu\alpha}g^{(\perp\perp)}_{\nu\nu'}g^{(\perp)}_{\alpha\alpha'}
U_j^{*\mu\nu'\alpha'}.
\end{eqnarray}
$d\Phi_n$ is the standard element of n-body phase space given by
\begin{equation}
d\Phi_n(p; p_1, \cdots p_n)=\delta^4(p-\sum^n_{i=1}p_i)
\prod^n_{i=1}\frac{d^3{\bf p}_i}{(2\pi)^32E_i}.
\end{equation}

\subsection*{Amplitudes for the doubly radiative
decay $\psi \rightarrow \gamma\gamma V(\rho, \omega, \phi)$}

This is a three step process: $\psi\to\gamma R$ with $R\to \gamma
V(\rho, \omega, \phi)$ and $\rho \to \pi^+\pi^-$,
$\omega\to\pi^0\pi^+\pi^-$, $\phi\to K^+ K^-$, here we number
$\pi^0$, $\pi^+$, $\pi^-$ as $0$, $1$, $2$. The intermediate
resonance state $X$ that may appear in the process with $J^{PC}$
values are $0^{++}, 0^{-+}$, $1^{++}$, $1^{-+}$, $2^{++}$,
$2^{-+}$, etc. Here J, P, C are the intrinsic spin, parity and
C-parity of the X particle, respectively.  For $\psi\to\gamma R$,
We denote the spin-orbital angular momenta between the photon and
$\psi$ by $S$ and $L$, respectively. The tensor describing the
first and the second steps will be denoted by $\tilde
T^{(L)}_{\mu_1\cdots\mu_L}$ and $\tilde
t^{(L_1)}_{\mu_1\cdots\mu_{L_1}}$,  respectively. The vector
describing the third step will be denoted by  $V_\mu$, where
$V(\rho,\phi)_\mu  = p_{1\mu} - p_{2\mu}$, here we use the
 fact that $\pi^+$ and $\pi^-$ (or $K^+$ and $K^-$) have equal masses; and
$$
V(\omega)_\mu = \epsilon^{\mu}_{\;\;\nu\lambda\sigma} p_1^\nu
p_2^\lambda p_0^\sigma [B_1(Q_{\omega\rho
0})f^{(\rho)}_{(12)}B_1(Q_{\rho 12}) + B_1(Q_{\omega\rho
2})f^{(\rho)}_{(01)}B_1(Q_{\rho 10}) + B_1(Q_{\omega\rho
1})f^{(\rho)}_{(02)}B_1(Q_{\rho 20})].
$$

Now we write the decay amplitude of the $\psi$  into two photons
and a vector in a general and compact form using the covariant
tensor formalism. There is one independent covariant tensor
amplitude
 for $\psi \rightarrow
\gamma 0^{++} \rightarrow \gamma \gamma V(\rho, \omega, \phi)$
\begin{equation}\label{0pp}
 U^{\mu\nu\alpha} = g^{\mu\nu}
V^\alpha f^{(R)} f^{(V)},
\end{equation}
where $f^{(V)}$ either $f^{(\rho,\phi)}_{(12)}$ or
$f^{(\omega)}_{(012)}$.

There is also one independent covariant tensor amplitude  for
$\psi \rightarrow \gamma 0^{-+} \rightarrow \gamma \gamma V(\rho,
\omega, \phi)$
\begin{equation}\label{0mp}
U^{\mu\nu\alpha} = \epsilon^{\mu\nu\lambda\sigma}p_\lambda \tilde
T^{(1)}_\sigma \epsilon^{\alpha\beta\rho\delta}K_\rho
t^{(1)}_{1\beta} V_\delta f^{(R)} f^{(V)}.
\end{equation}

For the production reaction  $\psi \rightarrow \gamma 1^{++}$
there are two  independent covariant tensor amplitudes ; there are
also two amplitudes  for the decay reaction $1^{++} \rightarrow
\gamma V(\rho, \omega, \phi)$, all in all we have four amplitudes
\begin{eqnarray}\label{1pp1}
U^{\mu\nu\alpha}_ 1\!&=&\! \epsilon^{\mu\nu\lambda\sigma}p_\lambda
\epsilon^{\alpha\beta\rho}_{\;\;\;\;\;\;\sigma}K_\rho V_\beta
f^{(R)} f^{(V)},
\\\label{1pp2}
 U^{\mu\nu\alpha}_ 2\!&=&\!\epsilon^{\mu\nu\lambda\sigma}p_\lambda
\tilde T^{(2)}_{\sigma\gamma}
 \epsilon^{\alpha\beta\rho\delta}K_\rho
\tilde t^{(2)\gamma}_\delta V_\beta f^{(R)} f^{(V)}
,\\\label{1pp3}
 U^{\mu\nu\alpha}_3\!&=&\! \epsilon^{\mu\nu\lambda\sigma}p_\lambda
 \epsilon^{\alpha\beta\rho\delta}
K_\rho \tilde t^{(2)}_{\sigma\delta} V_\beta f^{(R)}
f^{(V)},\\\label{1pp4} U^{\mu\nu\alpha}_4\!&=&\!
\epsilon^{\mu\nu\lambda\sigma}p_\lambda \tilde
T^{(2)}_{\sigma\delta}
 \epsilon^{\alpha\beta\rho\delta}
K_\rho  V_\beta f^{(R)} f^{(V)}.
\end{eqnarray}

For the production reaction  $\psi \rightarrow \gamma 1^{-+}$
there are two  independent covariant tensor amplitudes ; there are
also two amplitudes  for the decay reaction $1^{-+} \rightarrow
\gamma V(\rho, \omega, \phi)$, all in all we have four amplitudes
\begin{eqnarray}\label{1mp1}
U^{\mu\nu\alpha}_1 \!&=&\! g^{\mu\nu}\tilde T^{(1)}_\beta \tilde
t^{(1)\beta} V^\alpha f^{(R)} f^{(V)} ,
\\\label{1mp2}
U^{\mu\nu\alpha}_2 \!&=&\!\tilde T^{(1)\mu} \tilde t^{(1)\nu}
V^\alpha f^{(R)} f^{(V)} ,\\\label{1mp3}
U^{\mu\nu\alpha}_3\!&=&\!g^{\mu\nu}
 \tilde T^{(1)\alpha} \tilde
t^{(1)\beta}V_\beta f^{(R)} f^{(V)},\\\label{1mp4}
U^{\mu\nu\alpha}_4\!&=&\! \tilde T^{(1)\mu } g^{\nu\alpha} \tilde
t^{(1)\beta} V_\beta f^{(R)} f^{(V)}.
\end{eqnarray}

For the production reaction  $\psi \rightarrow \gamma 2^{++}$
there are three  independent covariant tensor amplitudes  ; there
are also three amplitudes  for the decay reaction $2^{++}
\rightarrow \gamma V(\rho, \omega, \phi)$, all in all we have nine
amplitudes
\begin{eqnarray}\label{2pp1}
U^{\mu\nu\alpha}_ 1 \!&=&\! P^{(2)\mu\nu\alpha\beta}(K) V_\beta
f^{(R)} f^{(V)},\\\label{2pp2}
 U^{\mu\nu\alpha}_ 2\!&=&\!
g^{\mu\nu} P^{(2)\lambda\sigma\rho\delta}(K)\tilde
T^{(2)}_{\lambda\sigma} \tilde t^{(2)}_{\rho\delta} V^\alpha
 f^{(R)} f^{(V)},\\\label{2pp3}
 U^{\mu\nu\alpha}_3
\!&=&\! P^{(2)\nu\sigma\alpha\lambda}(K) \tilde
T^{(2)\mu}_{\sigma} \tilde t^{(2)}_{\lambda\beta} V^\beta f^{(R)}
f^{(V)},\\\label{2pp4}
 U^{\mu\nu\alpha}_4 \!&=&\!P^{(2)\mu\nu\lambda\sigma}(K) \tilde
t^{(2)}_{\lambda\sigma} V^\alpha  f^{(R)} f^{(V)},\\\label{2pp5}
 U^{\mu\nu\alpha}_5 \!&=&\!P^{(2)\mu\nu\alpha\lambda}(K)
\tilde t^{(2)}_{\beta\lambda} V^\beta f^{(R)}
f^{(V)},\\\label{2pp6} U^{\mu\nu\alpha}_6 \!&=&\!
g^{\mu\nu}P^{(2)\lambda\sigma\alpha\beta}(K)\tilde
T^{(2)}_{\lambda\sigma} V_\beta  f^{(R)} f^{(V)},\\\label{2pp7}
U^{\mu\nu\alpha}_7 \!&=&\!g^{\mu\nu}
P^{(2)\lambda\sigma\alpha\delta}(K) \tilde
T^{(2)}_{\lambda\sigma}\tilde t^{(2)}_{\beta\delta} V^\beta
 f^{(R)} f^{(V)},\\\label{2pp8} U^{\mu\nu\alpha}_8
\!&=&\!  P^{(2)\nu\lambda\alpha\beta}(K) \tilde
T^{(2)\mu}_{\lambda }V_\beta  f^{(R)} f^{(V)},\\\label{2pp9}
U^{\mu\nu\alpha}_9 \!&=&\!  P^{(2)\nu\delta\lambda\sigma}(K)
\tilde T^{(2)\mu}_\delta \tilde t^{(2)}_{\lambda\sigma} V^\alpha
 f^{(R)} f^{(V)}.
\end{eqnarray}

For the production reaction  $\psi \rightarrow \gamma 2^{-+}$
there are three  independent covariant tensor amplitudes ; there
are also three amplitudes  for the decay reaction $2^{-+}
\rightarrow \gamma V(\rho, \omega, \phi)$, all in all we have nine
amplitudes
\begin{eqnarray}\label{2mp1}
U^{\mu\nu\alpha}_1 \!&=&\! \epsilon^{\mu\nu\lambda\sigma}p_\sigma
\tilde T^{(1)\gamma} \epsilon^{\alpha\beta\rho\xi}K_\xi \tilde
t^{(1)\delta}P^{(2)}_{\lambda\gamma\rho\delta}(K) V_\beta f^{(R)}
f^{(V)},
\\ \label{2mp2}
U^{\mu\nu\alpha}_2 \!&=&\! \epsilon^{\mu\nu\lambda\sigma}p_\sigma
\tilde T^{(3)}_{\lambda\gamma\delta}
\epsilon^{\alpha\beta\rho\xi}K_\xi \tilde
t^{(3)}_{\rho\gamma'\delta'} P^{(2)\gamma\delta\gamma'\delta'}(K)
V_\beta f^{(R)}f^{(V)} ,\\\label{2mp3}
 U^{\mu\nu\alpha}_ 3 \!&=&\!  \epsilon^{\nu\lambda\sigma\gamma}p_\gamma \tilde
T^{(3)\mu\lambda'}_\sigma \epsilon^{\beta\rho\delta\xi} K_\xi
\tilde t^{(3)\alpha\rho'}_{\delta}
P^{(2)}_{\lambda\lambda'\rho\rho'}(K) V_\beta f^{(R)} f^{(V)},
\\ \label{2mp4}
U^{\mu\nu\alpha}_ 4 \!&=&\!\epsilon^{\mu\nu\lambda\sigma}p_\sigma
\tilde T^{(1)\gamma} \epsilon^{\alpha\beta\rho\xi} K_\xi \tilde
t^{(3)\delta\zeta}_\rho P^{(2)}_{\lambda\gamma\delta\zeta }(K)
V_\beta f^{(R)} f^{(V)} ,\\
\label{2mp5} U^{\mu\nu\alpha}_5
\!&=&\!\epsilon^{\mu\nu\lambda\sigma}p_\sigma\tilde T^{(1)\gamma}
\epsilon^{\beta\rho\delta\xi} K_\xi
P^{(2)}_{\lambda\gamma\rho\zeta}(K)\tilde
t^{(3)\alpha\zeta}_{\delta} V_\beta f^{(R)} f^{(V)},
\\ \label{2mp6}
U^{\mu\nu\alpha}_6 \!&=&\! \epsilon^{\mu\nu\lambda\sigma}p_\sigma
\tilde T^{(3)\gamma\delta}_\lambda \epsilon^{\alpha\beta\rho\xi}
K_\xi \tilde t^{(1)\zeta}P^{(2)}_{\gamma\delta\rho\zeta}(K)
V^\beta f^{(R)} f^{(V)},\\
\label{1pp7} U^{\mu\nu\alpha}_7 \!&=&\!
\epsilon^{\mu\nu\lambda\sigma}p_\sigma \tilde
T^{(3)\gamma\delta}_\lambda
 \epsilon^{\beta\tau\rho\xi} K_\xi
\tilde t^{(3)\alpha\delta'}_\rho
P^{(2)}_{\gamma\delta\tau\delta'}(K) V_\beta f^{(R)} f^{(V)},\\
\label{2mp8} U^{\mu\nu\alpha}_ 8 \!&=&\!
\epsilon^{\nu\lambda\sigma\gamma} p_\gamma \tilde
T^{(3)\mu\zeta}_\sigma \epsilon^{\alpha\beta\rho\xi} K_\xi \tilde
t^{(1)\delta} P^{(2)}_{\lambda\zeta\rho\delta}(K)
  V_\beta f^{(R)} f^{(V)},\\
\label{2mp9} U^{\mu\nu\alpha}_ 9 \!&=&\!
\epsilon^{\nu\lambda\sigma\gamma} p_\gamma \tilde
T^{(3)\mu\delta}_\sigma \epsilon^{\alpha\beta\rho\xi} K_\xi \tilde
t^{(3)\lambda'\delta'}_\rho
  P^{(2)}_{\lambda\delta\lambda'\delta'}(K) V_\beta f^{(R)} f^{(V)}.
\end{eqnarray}

\section{Formalism for $\psi(2S)\to\gamma\chi_{cJ}$ with $\chi_{cJ}\to  K\bar K
\pi^+\pi^- $ and $2\pi^+2\pi^-$}

By following  Ref.\cite{zou} we denote the $\psi(2s)$ polarization
four-vector by $\psi_\mu(p,m_J)$ and  the photon polarization
vector by $e_\nu(q,m_\gamma)$. Then the general form for the decay
amplitude is
\begin{equation}
A=\psi_\mu(p,m_J) e^*_\nu(q,m_\gamma) A^{\mu\nu} =\psi_\mu(p,m_J)
e^*_\nu(q,m_\gamma)\sum_i\Lambda_i U_i^{\mu\nu}.
\end{equation}
The radiative decay cross section is given in :
\begin{eqnarray}
\frac{d\sigma}{d\Phi_n}\!&=&\!\frac{1}{2}\sum^2_{m_J=1}\sum^2_{m_\gamma=1}
\psi_{\mu}(p,m_J) e^*_{\nu}(q,m_\gamma)A^{\mu\nu}
\psi^*_{\mu'}(p,m_J) e_{\nu'}(q,m_\gamma)A^{*\mu'\nu'} \nonumber\\
\!&=&\!-\frac{1}{2}\sum^2_{m_J=1}
\psi_{\mu}(p,m_J)\psi^*_{\mu'}(p,m_J) g^{(\perp\perp)}_{\nu\nu'}
A^{\mu\nu}A^{*\mu'\nu'}\nonumber\\
&=& -\frac{1}{2}\sum^2_{\mu=1}A_{\mu\nu}
g^{(\perp\perp)}_{\nu\nu'}A^{*\mu\nu'}\nonumber\\
&=& -\frac{1}{2}\sum_{i,j}\Lambda_i\Lambda_j^*\sum^2_{\mu=1}
U_i^{\mu\nu}g^{(\perp\perp)}_{\nu\nu'}U_j^{*\mu\nu'}
 \equiv \sum_{i,j}P_{ij}\cdot F_{ij}
\end{eqnarray}
where $g^{(\perp\perp)}_{\nu\nu'}$ is given in $(\ref{photon})$
and
\begin{eqnarray}
P_{ij} &= P^*_{ji} &= \Lambda_i\Lambda^*_j, \\
F_{ij} &= F^*_{ji} &= -\frac{1}{2}\sum^2_{\mu=1}
U_i^{\mu\nu}g^{(\perp\perp)}_{\nu\nu'}U_j^{*\mu\nu'}.
\end{eqnarray}

Note that due to the special properties (massless and gauge
invariance) of the photon, the number of independent partial wave
amplitudes for a $\psi(2s)$ radiative decay is smaller than for
the corresponding decay to a massive vector meson \cite{zou}. We
come now to specific examples of reactions.

\subsection{$\psi\to\gamma\chi_{c0}\to\gamma K^+K^-\pi^+\pi^-$ }

We construct the covariant amplitudes $U_{\mu\nu}^i$ for this
channel. Here we number $K^+$, $K^-$, $\pi^+$, $\pi^-$ as 1, 2, 3,
4.
\begin{eqnarray}
<K^*_0 \bar K^*_0|1> \!&=&\! g_{\mu\nu}f_{(14)}^{(K^*_0)}f_{(23)}^{(\bar K^*_0)} , \\
 <K^*_0 \bar K^*_2|1> \!&=&\! g_{\mu\nu} \tilde
T_{[K_0 \bar K_2]}^{(2)\alpha\beta} \tilde
t_{(23)\alpha\beta}^{(2)}f_{(14)}^{(K^*_0)}f_{(23)}^{(\bar
K^*_2)}+\{1\leftrightarrow 2,3\leftrightarrow 4\} , \\
 <K^*_2 \bar K^*_2|1> \!&=&\! g_{\mu\nu}
\tilde t_{(14)}^{(2)\alpha\beta}\tilde t_{(23)\alpha\beta}^{(2)}
 f_{(14)}^{(K^*_2)}f_{(23)}^{(\bar K^*_2)} ,\\
<K^*_2 \bar K^*_2|2> \!&=&\! g_{\mu\nu} \tilde T_{[K_2 \bar
K_2]}^{(2)\alpha\beta} \tilde t_{(14)\alpha}^{(2)\gamma} \tilde
t_{(23)\beta\gamma}^{(2)}
 f_{(14)}^{(K^*_2)}f_{(23)}^{(\bar K^*_2)},\\
<K^* \bar K^*|1> \!&=&\! g_{\mu\nu}
\tilde t_{(14)}^{(1)\alpha} \tilde t_{(23) \alpha}^{(1)} f_{(14)}^{(K^*)}f_{(23)}^{(\bar K^*)}, \\
<K^* \bar K^*|2> \!&=&\! g_{\mu\nu}\tilde T_{[K^* \bar
K^*]}^{(2)\alpha\beta} \tilde t_{(14)\alpha}^{(1)} \tilde t_{(23)
\beta}^{(1)}
f_{(14)}^{(K^*)}f_{(23)}^{(\bar K^*)},\\
<K' K|K\rho> \!&=&\! g_{\mu\nu} \tilde T_{[K\rho]}^{(1)\alpha}
\tilde t_{(34)\alpha}^{(1)}f_{(134)}^{(K')}f_{(34)}^{(\rho)}
+ \{1\leftrightarrow 2,3\leftrightarrow 4\},\\
<K' K|K^*\pi> \!&=&\! g_{\mu\nu} \tilde T_{[K^*3]}^{(1)\alpha}
\tilde t_{(14)\alpha}^{(1)}f_{(134)}^{(K')}f_{(14)}^{(K^*)}
+ \{1\leftrightarrow 2,3\leftrightarrow 4\},\\
<K' K|K_0^*\pi> \!&=&\! g_{\mu\nu}
f_{(134)}^{(K')}f_{(14)}^{(K_0^*)}
+\{1\leftrightarrow 2,3\leftrightarrow 4\},\\
<K^*_1 K|K\rho> \!&=&\! g_{\mu\nu} \tilde g_{\alpha\beta}
(p_{K_1^*}) \tilde T_{[K^*_1\bar K]}^{(1)\alpha} \tilde
t_{(34)}^{(1)\beta}f_{(134)}^{(K_1^*)}f_{(34)}^{(\rho)}
+ \{1\leftrightarrow 2,3\leftrightarrow 4\},\\
<K^*_1 K|K^*\pi>_1 \!&=&\! g_{\mu\nu} \tilde g_{\alpha\beta}
(p_{K_1^*}) \tilde T_{[K^*_1\bar K]}^{(1)\alpha} \tilde
t_{(14)}^{(1)\beta}f_{(134)}^{(K_1^*)}f_{(14)}^{(K^*)}
+ \{1\leftrightarrow 2,3\leftrightarrow 4\},\\
<K^*_1 K|K^*\pi>_2 \!&=&\! g_{\mu\nu} \tilde g_{\alpha\beta}
(p_{K_1^*}) \tilde T_{[K^*_1\bar K]}^{(1)\alpha} \tilde
t^{(2)\beta\sigma}_{(K^* \pi)} \tilde t_{(14)\sigma}^{(1)}
f_{(134)}^{(K_1^*)}f_{(14)}^{(K^*)}+
\{1\leftrightarrow 2,3\leftrightarrow 4\},\\
<K^*_1 K|K_0^*\pi> \!&=&\! g_{\mu\nu} \tilde g_{\alpha\beta}
(p_{K_1^*}) \tilde T_{[K^*_1\bar K]}^{(1)\alpha} \tilde t_{(K_0^*
\pi)}^{(1)\beta} f_{(134)}^{(K_1^*)}f_{(14)}^{(K_0^*)}
+\{1\leftrightarrow 2,3\leftrightarrow 4\},\\
<K_2 K|K^*\pi>\!&=&\! g_{\mu\nu}
P^{(2)}_{\alpha\beta\lambda\sigma}(p_{K_2})
 \tilde T_{[K_2\bar K] }^{(2)\alpha\beta} \tilde
t_{(K^* \pi)}^{(1)\sigma} \tilde
t^{(1)\lambda}_{(14)}f_{(134)}^{(K_1^*)}
f_{(14)}^{(K^*)} + \{1\leftrightarrow 2,3\leftrightarrow 4\},\\
<f_0f'_0|1> \!&=&\! g_{\mu\nu}f_{(34)}^{(f_0)}f_{(12)}^{(f'_0)} , \\
<f_0f_2|1> \!&=&\! g_{\mu\nu} \tilde T_{[f_0
f_2]}^{(2)\alpha\beta} \tilde
t_{(34)\alpha\beta}^{(2)}f_{(34)}^{(f_0)}f_{(12)}^{(f_2)},\\
<f_2f'_2|1> \!&=&\! g_{\mu\nu} \tilde
t_{(12)}^{(2)\alpha\beta}\tilde
t_{(34)\alpha\beta}^{(2)}f_{(12)}^{(f_2)}f_{(34)}^{(f'_2)}
, \\
<f_2f'_2|2> \!&=&\! g_{\mu\nu} \tilde T_{[f_2
f'_2]}^{(2)\alpha\beta} \tilde t_{(12)\alpha}^{(2)\gamma} \tilde
t_{(34)\beta\gamma}^{(2)}f_{(12)}^{(f_2)}f_{(34)}^{(f'_2)}.
\end{eqnarray}

\subsection{$\psi\to\gamma\chi_{c1}\to\gamma K^+K^-\pi^+\pi^-$ }

In this subsection we construct the amplitudes $U_{\mu\nu}^i$ for
the process $\psi\to\gamma\chi_{c1}\to\gamma K^+ K^-\pi^+\pi^-$.
The most likely intermediate states are: $K^*_0 \bar K^*$, $K^*_0
\bar K^*_2$, $K^*_2 \bar K^*$, $K_2^* \bar K_2^*$, $K^* \bar K^*$,
$K_1^* K$, $K_2^* K$ with $K_0^*$, $K_2^*$, $K^* \to K\pi$, $K_1^*
\to \rho K$, $K^*\pi$, $K_0^*\pi$,  and  $f_0 f_2$, $f_2 f'_2$
with $f_0 \to \pi^+\pi^-$, $f'_0 \to K^+ K^-$, $f_2 \to K^+ K^-$
and $f'_2 \to \pi^+\pi^-$.

\begin{eqnarray}
<K^*_0 \bar K^*|1>
\!&=&\!\varepsilon_{\mu\nu\alpha\beta}p^\beta_{\psi}
\varepsilon_{\lambda\gamma\delta\xi} p^\xi_{\chi_{c1}}
g^{\alpha\delta} [\tilde t^{(1)\gamma}_{[K_0^* \bar K^*]}\tilde
t^{(1)\lambda}_{(23)} f_{(14)}^{(K^*_0)}f_{(23)}^{(\bar
K^*)} \nonumber\\&&+\{1\leftrightarrow 2,3\leftrightarrow 4\}],\\
<K^*_0 \bar K^*|2>
\!&=&\!\varepsilon_{\rho\nu\alpha\beta}p^\beta_{\psi}
\varepsilon_{\lambda\gamma\delta\xi} p^\xi_{\chi_{c1}} q^\rho
q_\mu g^{\alpha\delta} [\tilde t^{(1)\gamma}_{[K_0^* \bar
K^*]}\tilde t^{(1)\lambda}_{(23)}f_{(14)}^{(K^*_0)}f_{(23)}^{(\bar
K^*)}
\nonumber\\&&+\{1\leftrightarrow 2,3\leftrightarrow 4\}],\\
<K^*_0 \bar K^*_2|1> \!&=&\! \varepsilon_{\mu\nu\alpha\beta}
p^\beta_{\psi} \varepsilon_{\lambda\gamma\delta\xi}
p^\xi_{\chi_{c1}}  g^{\alpha\delta} [\tilde T^{(2)\gamma}_{[K^*_0
\bar K^*_2] \sigma} \tilde t_{(23)}^{(2)\lambda\sigma}
f_{(14)}^{(K^*_0)}f_{(23)}^{(\bar K^*_2)}\nonumber\\&&+\{1\leftrightarrow 2,3\leftrightarrow 4\}],\\
<K^*_0 \bar K^*_2|2> \!&=&\! \varepsilon_{\rho\nu\alpha\beta}
p^\beta_{\psi} q^\rho q_\mu \varepsilon_{\lambda\gamma\delta\xi}
p^\xi_{\chi_{c1}} g^{\alpha\delta}
 [\tilde T^{(2)\gamma}_{[K^*_0 \bar K^*_2] \sigma} \tilde t_{(23)}^{(2)\lambda\sigma}
f_{(14)}^{(K^*_0)}f_{(23)}^{(\bar K^*_2)}\nonumber\\&&+\{1\leftrightarrow 2,3\leftrightarrow 4\}],\\
<K^*_2 \bar K^*|1>
\!&=&\!\varepsilon_{\mu\nu\alpha\beta}p^\beta_{\psi}
\varepsilon_{\lambda\gamma\delta\xi} p^\xi_{\chi_{c1}}
g^{\alpha\delta} [\tilde t^{(1)\gamma}_{[K_2^* \bar K^*]} \tilde
t^{(2)\lambda\sigma}_{(14)} \tilde t^{(1)}_{(23)\sigma}
f_{(14)}^{(K^*_2)}f_{(23)}^{(\bar K^*)} \nonumber\\&&+\{1\leftrightarrow 2,3\leftrightarrow 4\}],\\
<K^*_2 \bar K^*|2>
\!&=&\!\varepsilon_{\mu\nu\alpha\beta}p^\beta_{\psi}
\varepsilon_{\lambda\gamma\delta\xi} p^\xi_{\chi_{c1}}
g^{\alpha\delta} [\tilde t^{(1)}_{[K_2^* \bar K^*]\sigma} \tilde
t^{(2)\lambda\sigma}_{(14)} \tilde t^{(1)\gamma}_{(23)}
f_{(14)}^{(K^*_2)}f_{(23)}^{(\bar K^*)} \nonumber\\&&+\{1\leftrightarrow 2,3\leftrightarrow 4\}],\\
<K^*_2 \bar K^*|3> \!&=&\!\varepsilon_{\rho\nu\alpha\beta}
p^\beta_{\psi} q^\rho q_\mu  \varepsilon_{\lambda\gamma\delta\xi}
p^\xi_{\chi_{c1}} g^{\alpha\delta} [\tilde t^{(1)\gamma}_{[K_2^*
\bar K^*]} \tilde t^{(2)\lambda\sigma}_{(14)} \tilde
t^{(1)}_{(23)\sigma}
f_{(14)}^{(K^*_2)}f_{(23)}^{(\bar K^*)} \nonumber\\&&+\{1\leftrightarrow 2,3\leftrightarrow 4\}],\\
<K^*_2 \bar K^*|4> \!&=&\! \varepsilon_{\rho\nu\alpha\beta}
p^\beta_{\psi} q^\rho q_\mu \varepsilon_{\lambda\gamma\delta\xi}
p^\xi_{\chi_{c1}} g^{\alpha\delta} [\tilde t^{(1)}_{[K_2^* \bar
K^*]\sigma} \tilde t^{(2)\lambda\sigma}_{(14)} \tilde
t^{(1)\gamma}_{(23)}
f_{(14)}^{(K^*_2)}f_{(23)}^{(\bar K^*)} \nonumber\\&&+\{1\leftrightarrow 2,3\leftrightarrow 4\}],\\
<K^*_2 \bar K^*_2|1> \!&=&\! \varepsilon_{\mu\nu\alpha\beta}
p^\beta_{\psi} \varepsilon_{\lambda\gamma\delta\xi}
p^\xi_{\chi_{c1}}  g^{\alpha\delta}
 \tilde t_{(14)}^{(2)\lambda\sigma} \tilde t^{(2)\gamma}_{(23)\sigma}
f_{(14)}^{(K^*_2)}f_{(23)}^{(\bar K^*_2)},\\
<K^*_2 \bar K^*_2|2> \!&=&\! \varepsilon_{\rho\nu\alpha\beta}
p^\beta_{\psi} q^\rho q_\mu \varepsilon_{\lambda\gamma\delta\xi}
p^\xi_{\chi_{c1}}  g^{\alpha\delta}
 \tilde t_{(14)}^{(2)\lambda\sigma} \tilde t^{(2)\gamma}_{(23)\sigma}
f_{(14)}^{(K^*_2)}f_{(23)}^{(\bar K^*_2)},\\
<K^* \bar K^*|1> \!&=&\! \varepsilon_{\mu\nu\alpha\beta}
p^\beta_{\psi} \varepsilon_{\lambda\sigma\gamma\delta}
p^\delta_{\chi_{c1}}
   g^{\alpha\gamma}
 \tilde t^{(1)\lambda}_{(14)} \tilde t^{(1)\sigma}_{(23)}
f_{(14)}^{(K^*)}f_{(23)}^{(\bar K^*)},\\
<K^* \bar K^*|2> \!&=&\! \varepsilon_{\rho\nu\alpha\beta}
p^\beta_{\psi} q^\rho q_\mu
\varepsilon_{\lambda\sigma\gamma\delta} p^\delta_{\chi_{c1}}
   g^{\alpha\gamma}
 \tilde t^{(1)\lambda}_{(14)} \tilde t^{(1)\sigma}_{(23)}
f_{(14)}^{(K^*)}f_{(23)}^{(\bar K^*)},\\
<K^*_1  K|K \rho>_1 \!&=&\! \varepsilon_{\mu\nu\alpha\beta}
p^\beta_{\psi} \varepsilon_{\lambda\sigma\gamma\delta}
p_{\chi_{c1}}^\delta
   g^{\alpha\gamma}
\tilde T^{(1)\sigma}_{[K^*_1  K]} \tilde g^{\lambda\xi}(p_{K^*_1})
\nonumber\\&&
 [\tilde t^{(1)}_{(34)\xi}
f_{(134)}^{(K^*_1)}f_{(34)}^{(\rho)} +
\{1\leftrightarrow 2,3\leftrightarrow 4\}],\\
<K^*_1  K|K \rho>_2 \!&=&\! \varepsilon_{\eta\nu\alpha\beta}
p^\beta_{\psi} q^\eta q_\mu
\varepsilon_{\lambda\sigma\gamma\delta} p_{\chi_{c1}}^\delta
   g^{\alpha\gamma}
\tilde T^{(1)\sigma}_{[K^*_1  K]} \tilde g^{\lambda\xi}(p_{K^*_1})
\nonumber\\&&
 [\tilde t^{(1)}_{(34)\xi}
f_{(134)}^{(K^*_1)}f_{(34)}^{(\rho)} +
\{1\leftrightarrow 2,3\leftrightarrow 4\}],\\
<K^*_1  K|K^* \pi>_1 \!&=&\! \varepsilon_{\mu\nu\alpha\beta}
p^\beta_{\psi} \varepsilon_{\lambda\sigma\gamma\delta}
p_{\chi_{c1}}^\delta
   g^{\alpha\gamma}
\tilde T^{(1)\sigma}_{[K^*_1  K]} \tilde g^{\lambda\xi}(p_{K^*_1})
\nonumber\\&&
 [\tilde t^{(1)}_{(14)\xi}
f^{(K^*_1)}_{(134)} f^{(K^*)}_{(14)} +
\{1\leftrightarrow 2,3\leftrightarrow 4\}],\\
<K^*_1  K|K^* \pi>_2 \!&=&\! \varepsilon_{\rho\nu\alpha\beta}
p^\beta_{\psi} q^\rho q_\mu
\varepsilon_{\lambda\sigma\gamma\delta} p_{\chi_{c1}}^\delta
   g^{\alpha\gamma}
\tilde T^{(1)\sigma}_{[K^*_1  K]} \tilde g^{\lambda\xi}(p_{K^*_1})
\nonumber\\&&
 [\tilde t^{(1)}_{(14)\xi}
f^{(K^*_1)}_{(134)} f^{(K^*)}_{(14)} +
\{1\leftrightarrow 2,3\leftrightarrow 4\}],\\
<K^*_1  K|K^*_0 \pi>_1 \!&=&\! \varepsilon_{\mu\nu\alpha\beta}
p^\beta_{\psi} \varepsilon_{\lambda\sigma\gamma\delta}
p_{\chi_{c1}}^\delta
  g^{\alpha\gamma}
\tilde T^{(1)\sigma}_{[K^*_1  K]} \tilde g^{\lambda\xi}(p_{K^*_1})
\nonumber\\&& \tilde t^{(1)}_{[K^*_0 \pi]\xi}
 [f^{(K^*_1)}_{(134)} f^{(K^*)}_{(14)} +
\{1\leftrightarrow 2,3\leftrightarrow 4\}],\\
<K^*_1  K|K^*_0 \pi>_2 \!&=&\! \varepsilon_{\rho\nu\alpha\beta}
p^\beta_{\psi} q^\rho q_\mu
\varepsilon_{\lambda\sigma\gamma\delta} p_{\chi_{c1}}^\delta
  g^{\alpha\gamma}
\tilde T^{(1)\sigma}_{[K^*_1  K]} \tilde g^{\lambda\xi}(p_{K^*_1})
\nonumber\\&& \tilde t^{(1)}_{[K^*_0 \pi]\xi}
 [f^{(K^*_1)}_{(134)} f^{(K^*)}_{(14)} +
\{1\leftrightarrow 2,3\leftrightarrow 4\}],\\
<K^*_2  K|K^* \pi>_1 \!&=&\! \varepsilon_{\mu\nu\alpha\beta}
p^\beta_{\psi} \varepsilon^{\gamma\xi\delta\tau} p_{\chi_{c1}\tau}
   \tilde g^{\alpha\lambda}(p_{\chi_{c1}})
  P^{(2)}_{\lambda\sigma\gamma\xi}(p_{K^*_2})\nonumber\\&&
\tilde T^{(1)\sigma}_{[K^*_2  K]} \tilde T^{(2)}_{[K^*\pi]
\delta\eta}
 [\tilde t^{(1)\eta}_{(14)}f^{(K^*_1)}_{(134)} f^{(K^*)}_{(14)} +
\{1\leftrightarrow 2,3\leftrightarrow 4\}],\\
<K^*_2  K|K^* \pi>_2 \!&=&\! \varepsilon_{\rho\nu\alpha\beta}
p^\beta_{\psi} q^\rho q_\mu \varepsilon^{\gamma\xi\delta\tau}
p_{\chi_{c1}\tau}
   \tilde g^{\alpha\lambda}(p_{\chi_{c1}})
  P^{(2)}_{\lambda\sigma\gamma\xi}(p_{K^*_2})\nonumber\\&&
\tilde T^{(1)\sigma}_{[K^*_2  K]} \tilde T^{(2)}_{[K^*\pi]
\delta\eta}
 [\tilde t^{(1)\eta}_{(14)}f^{(K^*_1)}_{(134)} f^{(K^*)}_{(14)} +
\{1\leftrightarrow 2,3\leftrightarrow 4\}],\\
<f_0 f_2|1> \!&=&\!  \varepsilon_{\mu\nu\alpha\beta}
p^\beta_{\psi} \varepsilon_{\lambda\gamma\delta\xi}
p^\xi_{\chi_{c1}}
  g^{\alpha\delta}
 \tilde T^{(2)\gamma}_{[f_0 f_2] \sigma} \tilde t_{(12)}^{(2)\lambda\sigma}
f_{(34)}^{(f_0)}f_{(12)}^{(f_2)},\\
<f_0 f_2|2> \!&=&\!  \varepsilon_{\rho\nu\alpha\beta}
p^\beta_{\psi} q^\rho q_\mu \varepsilon_{\lambda\gamma\delta\xi}
p^\xi_{\chi_{c1}}
  g^{\alpha\delta}
 \tilde T^{(2)\gamma}_{[f_0 f_2] \sigma} \tilde t_{(12)}^{(2)\lambda\sigma}
f_{(34)}^{(f_0)}f_{(12)}^{(f_2)},\\
<f_2 \bar f'_2|1> \!&=&\! \varepsilon_{\mu\nu\alpha\beta}
p^\beta_{\psi} \varepsilon_{\lambda\gamma\delta\xi}
p_{\chi_{c1}}^\xi  g^{\alpha\delta}
 \tilde t_{(12)}^{(2)\lambda\sigma} \tilde t^{(2)\gamma}_{(34)\sigma}
f_{(12)}^{(f_2)} f_{(34)}^{(f'_2)},\\
<f_2 \bar f'_2|2> \!&=&\! \varepsilon_{\rho\nu\alpha\beta}
p^\beta_{\psi} q^\rho q_\mu \varepsilon_{\lambda\gamma\delta\xi}
p_{\chi_{c1}}^\xi  g^{\alpha\delta}
 \tilde t_{(12)}^{(2)\lambda\sigma} \tilde t^{(2)\gamma}_{(34)\sigma}
f_{(12)}^{(f_2)} f_{(34)}^{(f'_2)}.
\end{eqnarray}

\subsection{$\psi\to\gamma\chi_{c2}\to\gamma K^+K^-\pi^+\pi^-$ }

We construct the amplitudes $U_{\mu\nu}^i$ for the channel
$\psi\to\gamma\chi_{c2}\to\gamma K^+K^-\pi^+\pi^-$. The most
possible intermediate states are the same as for
$\psi\to\gamma\chi_{c1}\to\gamma K^+K^-\pi^+\pi^-$.
\begin{eqnarray}
<K_0^* \bar K_0^* |1> \!&=&\! g_{\mu\nu} \tilde
T^{(2)\alpha\beta}_{[\gamma\chi_{c2}] } \tilde T^{(2)}_{[K_0^*
\bar K_0^*]\alpha\beta}
 f_{(14)}^{(K_0^*)} f_{(23)}^{(\bar K_0^*)} , \\
<K_0^* \bar K_0^* |2> \!&=&\! \tilde T^{(2)}_{[K_0^* \bar K_0^*]
\mu\nu}
 f_{(14)}^{(K_0^*)} f_{(23)}^{(\bar K_0^*)} , \\
<K_0^* \bar K_0^* |3> \!&=&\! \tilde T^{(2)
\alpha}_{[\gamma\chi_{c2}] \mu} \tilde T^{(2) }_{[K_0^* \bar
K_0^*] \nu\alpha}
 f_{(14)}^{(K_0^*)} f_{(23)}^{(\bar K_0^*)} , \\
<K_0^* \bar K^* |1>
\!&=&\!P^{(2)}_{\mu\nu\lambda\sigma}(p_{\chi_{c2}})[ \tilde
t^{(1)\sigma}_{[K_0^*\bar K^*]} \tilde
t^{(1)\lambda}_{(23)}f_{(14)}^{(K_0^*)} f_{(23)}^{(\bar K^*)}
+ \{1\leftrightarrow 2,3\leftrightarrow 4\}] , \\
<K_0^* \bar K^* |2>
\!&=&\!P^{(2)}_{\beta\nu\lambda\sigma}(p_{\chi_{c2}})\tilde
T^{(2)\beta}_{[\gamma\chi_{c2}]\mu}[ \tilde
t^{(1)\sigma}_{[K_0^*\bar K^*]} \tilde
t^{(1)\lambda}_{(23)}f_{(14)}^{(K_0^*)} f_{(23)}^{(\bar K^*)}
+ \{1\leftrightarrow 2,3\leftrightarrow 4\}] , \\
<K_0^* \bar K^* |3> \!&=&\! g_{\mu\nu}
P^{(2)}_{\alpha\beta\lambda\sigma}(p_{\chi_{c2}})\tilde
T^{(2)\alpha\beta}_{[\gamma\chi_{c2}]}[ \tilde
t^{(1)\sigma}_{[K_0^*\bar K^*]} \tilde
t^{(1)\lambda}_{(23)}f_{(14)}^{(K_0^*)} f_{(23)}^{(\bar K^*)}
+ \{1\leftrightarrow 2,3\leftrightarrow 4\}] , \\
<K_0^* \bar K_2^* |1> \!&=&\! g_{\mu\nu} \tilde
T^{(2)\alpha\beta}_{[\gamma\chi_{c2}] }
 P^{(2)}_{\alpha\beta\lambda\sigma}(p_{\chi_{c2}})
\tilde t^{(2)\lambda\sigma}_{(23)}
 f_{(14)}^{(K_0^*)} f_{(23)}^{(\bar K_2^*)}+ \{1\leftrightarrow 2,3\leftrightarrow 4\}] , \\
<K_0^* \bar K_2^* |2> \!&=&\!
 P^{(2)}_{\mu\nu\lambda\sigma}(p_{\chi_{c2}})
\tilde t^{(2)\lambda\sigma}_{(23) }
 f_{(14)}^{(K_0^*)} f_{(23)}^{(\bar K_2^*)} + \{1\leftrightarrow 2,3\leftrightarrow 4\}], \\
 <K_0^* \bar K_2^* |3> \!&=&\! \tilde T^{(2)\alpha}_{[\gamma\chi_{c2}] \mu}
 P^{(2)}_{\nu\alpha\lambda\sigma}(p_{\chi_{c2}})
\tilde t^{(2) \lambda\sigma}_{(23)}
 f_{(14)}^{(K_0^*)} f_{(23)}^{(\bar K_2^*)} + \{1\leftrightarrow 2,3\leftrightarrow 4\}], \\
<K_2^* \bar K_2^* |1> \!&=&\! g_{\mu\nu} \tilde
T^{(2)\alpha\beta}_{[\gamma\chi_{c2}] }
  P^{(2)}_{\alpha\beta\lambda\sigma} (p_{\chi_{c2}})
\tilde t^{(2)\sigma\rho}_{(14)}
 \tilde t_{(23)\rho}^{(2)\lambda}
 f_{(14)}^{(K_2^*)} f_{(23)}^{(\bar K_2^*)},\\
<K_2^* \bar K_2^* |2> \!&=&\!
  P^{(2)}_{\mu\nu\lambda\sigma} (p_{\chi_{c2}})
\tilde t^{(2)\sigma\rho}_{(14)}
 \tilde t_{(23)\rho}^{(2)\lambda}
 f_{(14)}^{(K_2^*)} f_{(23)}^{(\bar K_2^*)},\\
<K_2^* \bar K_2^* |3> \!&=&\! \tilde
T^{(2)\alpha}_{[\gamma\chi_{c2}] \mu}
  P^{(2)}_{\nu\alpha\lambda\sigma} (p_{\chi_{c2}})\tilde t^{(2)\sigma\rho}_{(14)}
 \tilde t_{(23)\rho}^{(2)\lambda}
 f_{(14)}^{(K_2^*)} f_{(23)}^{(\bar K_2^*)},\\
 <K^* \bar K^*|1> \!&=&\!g_{\mu\nu} \tilde T^{(2)\alpha\beta}_{[\gamma\chi_{c2}] }
 P^{(2)}_{\alpha\beta\lambda\sigma} (p_{\chi_{c2}})
\tilde t^{(1)\lambda}_{(14)}
 \tilde t_{(23)}^{(1)\sigma} f_{(14)}^{(K^*)} f_{(23)}^{(\bar K^*)},\\
<K^* \bar K^*|2> \!&=&\!
 P^{(2)}_{\mu\nu\lambda\sigma} (p_{\chi_{c2}})
\tilde t^{(1)\lambda}_{(14)}
 \tilde t_{(23)}^{(1)\sigma} f_{(14)}^{(K^*)} f_{(23)}^{(\bar K^*)},\\
<K^* \bar K^*|3> \!&=&\! \tilde T^{(2)\alpha}_{[\gamma\chi_{c2}]
\mu}
 P^{(2)}_{\nu\alpha\lambda\sigma} (p_{\chi_{c2}})
\tilde t^{(1)\lambda}_{(14)}
 \tilde t_{(23)}^{(1)\sigma} f_{(14)}^{(K^*)} f_{(23)}^{(\bar K^*)},\\
 <K_1^* K|K \rho>_1 \!&=&\!g_{\mu\nu} \tilde T^{(2)\alpha\beta}_{[\gamma\chi_{c2}] }
 P^{(2)}_{\alpha\beta\lambda\sigma} (p_{\chi_{c2}})
 \tilde T^{(1)\sigma}_{[K_1^* \bar K]} \tilde g^{\lambda\delta}(p_{K_1^*})
 \nonumber\\&&
[\tilde t^{(1)}_{(34)\delta}
 f^{(K^*_1)}_{(134)} f_{(34)}^{(\rho)}   +
\{1\leftrightarrow 2,3\leftrightarrow 4\}],\\
<K_1^* K|K \rho>_2 \!&=&\!
 P^{(2)}_{\mu\nu\lambda\sigma} (p_{\chi_{c2}})
 \tilde T^{(1)\sigma}_{[K_1^* \bar K]} \tilde g^{\lambda\delta}(p_{K_1^*})
[\tilde t^{(1)}_{(34)\delta}
f^{(K^*_1)}_{(134)} f_{(34)}^{(\rho)}   +
\{1\leftrightarrow 2,3\leftrightarrow 4\}],\\
 <K_1^* K|K \rho>_3 \!&=&\! \tilde T^{(2)\alpha}_{[\gamma\chi_{c2}] \mu }
 P^{(2)}_{\nu\alpha\lambda\sigma} (p_{\chi_{c2}})
 \tilde T^{(1)\sigma}_{[K_1^* \bar K]} \tilde g^{\lambda\delta}(p_{K_1^*})
[\tilde t^{(1)}_{(34)\delta}
  f^{(K^*_1)}_{(134)} f_{(34)}^{(\rho)} +
\{1\leftrightarrow 2,3\leftrightarrow 4\}],\\
<K_1^* K|K^*\pi>_1 \!&=&\!g_{\mu\nu} \tilde T^{(2)\alpha\beta}_{[\gamma\chi_{c2}] }
 P^{(2)}_{\alpha\beta\lambda\sigma} (p_{\chi_{c2}})
 \tilde T^{(1)\sigma}_{[K_1^* \bar K]} \tilde g^{\lambda\delta}(p_{K_1^*})\nonumber\\&&
[\tilde t^{(1)}_{(14)\delta}
f^{(K^*_1)}_{(134)} f_{(14)}^{(K^*)}  + \{1\leftrightarrow 2,3 \leftrightarrow 4\}],\\
<K_1^* K|K^*\pi>_2 \!&=&\!
 P^{(2)}_{\mu\nu\lambda\sigma} (p_{\chi_{c2}})
 \tilde T^{(1)\sigma}_{[K_1^* \bar  K]} \tilde g^{\lambda\delta}(p_{K_1^*})
[\tilde t^{(1)}_{(14)\delta}
 f^{(K^*_1)}_{(134)} f_{(14)}^{(K^*)}  + \{1\leftrightarrow 2,3 \leftrightarrow 4\}],\\
<K_1^* K|K^*\pi>_3 \!&=&\! \tilde T^{(2)\alpha}_{[\gamma\chi_{c2}] \mu}
 P^{(2)}_{\nu\alpha\lambda\sigma} (p_{\chi_{c2}})
 \tilde T^{(1)\sigma}_{[K_1^* \bar K]} \tilde g^{\lambda\delta}(p_{K_1^*})\nonumber\\&&
[\tilde t^{(1)}_{(14)\delta}
  f^{(K^*_1)}_{(134)} f_{(14)}^{(K^*)} + \{1\leftrightarrow 2,3 \leftrightarrow 4\}] ,\\
<K_1^* K|K^*_0\pi>_1 \!&=&\!g_{\mu\nu} \tilde T^{(2)\alpha\beta}_{[\gamma\chi_{c2}] }
 P^{(2)}_{\alpha\beta\lambda\sigma} (p_{\chi_{c2}})
 \tilde T^{(1)\sigma}_{[K_1^* \bar K]} \tilde g^{\lambda\delta}(p_{K_1^*})
\tilde t^{(1)}_{(K_0^*\pi)\delta}\nonumber\\&&
 [f^{(K^*_1)}_{(134)} f_{(14)}^{(K^*_0)} + \{1\leftrightarrow 2,3 \leftrightarrow 4\}],\\
<K_1^* K|K^*_0\pi>_2 \!&=&\!
 P^{(2)}_{\mu\nu\lambda\sigma} (p_{\chi_{c2}})
 \tilde T^{(1)\sigma}_{[K_1^* \bar K]} \tilde g^{\lambda\delta}(p_{K_1^*})
\tilde t^{(1)}_{(K_0^*\pi)\delta}
[ f^{(K^*_1)}_{(134)} f_{(14)}^{(K^*_0)} + \{1\leftrightarrow 2,3 \leftrightarrow 4\}] ,\\
<K_1^* K|K^*_0\pi>_3 \!&=&\! \tilde T^{(2)\alpha}_{[\gamma\chi_{c2}] \mu}
 P^{(2)}_{\nu\alpha\lambda\sigma} (p_{\chi_{c2}})
 \tilde T^{(1)\sigma}_{[K_1^* \bar K]} \tilde g^{\lambda\delta}(p_{K_1^*})
\tilde t^{(1)}_{(K_0^*\pi)\delta}\nonumber\\&&
 [f^{(K^*_1)}_{(134)} f_{(14)}^{(K^*_0)} + \{1\leftrightarrow 2,3 \leftrightarrow 4\}] ,\\
<K_2^* K|K^*\pi>_1 \!&=&\!g_{\mu\nu}
\tilde T^{(2)\alpha\beta}_{[\gamma\chi_{c2}] }
 P^{(2)}_{\alpha\beta\lambda\xi} (p_{\chi_{c2}})
 P^{(2)\xi\sigma\xi'\sigma'}(p_{K_2^*})\varepsilon^{\lambda\gamma\delta}_\sigma p_{\chi_{c2}\delta}
\varepsilon^{\gamma'\eta'\delta'}_{\sigma'} p_{K^*_2 \delta'}
  \nonumber\\&&
 \tilde T^{(1)}_{[K_2^* \bar K]\gamma}
\tilde t^{(2)}_{(K^*\pi)\gamma'\xi'} [\tilde t^{(1)}_{(14)\eta'}
f^{(K^*_1)}_{(134)} f_{(14)}^{(K^*)} + \{1\leftrightarrow 2,3 \leftrightarrow 4\}],\\
<K_2^* K|K^*\pi>_2 \!&=&\!
 P^{(2)}_{\mu\nu\lambda\xi} (p_{\chi_{c2}})
 P^{(2)\xi\sigma\xi'\sigma'}(p_{K_2^*})\varepsilon^{\lambda\gamma\delta}_\sigma p_{\chi_{c2}\delta}
\varepsilon^{\gamma'\eta'\delta'}_{\sigma'} p_{K^*_2 \delta'}
  \nonumber\\&&
 \tilde T^{(1)}_{[K_2^* \bar K]\gamma}
\tilde t^{(2)}_{(K^*\pi)\gamma'\xi'}[ \tilde t^{(1)}_{(14)\eta'}
f^{(K^*_1)}_{(134)} f_{(14)}^{(K^*)} + \{1\leftrightarrow 2,3 \leftrightarrow 4\}],\\
<K_2^* K|K^*\pi>_3 \!&=&\!
\tilde T^{(2)\alpha}_{[\gamma\chi_{c2}] \mu}
 P^{(2)}_{\nu\alpha\lambda\xi} (p_{\chi_{c2}})
 P^{(2)\xi\sigma\xi'\sigma'}(p_{K_2^*}) \varepsilon^{\lambda\gamma\delta}_\sigma p_{\chi_{c2}\delta}
\varepsilon^{\gamma'\eta'\delta'}_{\sigma'} p_{K^*_2 \delta'}
  \nonumber\\&&
 \tilde T^{(1)}_{[K_2^* \bar K]\gamma}
\tilde t^{(2)}_{(K^*\pi)\gamma'\xi'} [\tilde t^{(1)}_{(14)\eta'}
f^{(K^*_1)}_{(134)} f_{(14)}^{(K^*)} + \{1\leftrightarrow 2,3 \leftrightarrow 4\}],\\
 <f_0 f'_0|1>\!&=&\! g_{\mu\nu}
\tilde T^{(2)\alpha\beta}_{[\gamma\chi_{c2}] }
  \tilde T^{(2)}_{[f_0 f'_0]\alpha\beta}
f_{(12)}^{(f'_0)} f_{(34)}^{(f_0)},\\
<f_0 f'_0|2>\!&=&\!
  \tilde T^{(2)}_{[f_0 f'_0]\mu\nu}
f_{(12)}^{(f'_0)} f_{(34)}^{(f_0)},\\
<f_0 f'_0|3>\!&=&\!
\tilde T^{(2)\alpha}_{[\gamma\chi_{c2}] \mu}
  \tilde T^{(2)}_{[f_0 f'_0]\nu\alpha}
f_{(12)}^{(f'_0)} f_{(34)}^{(f_0)},\\
<f_0 f_2|1>\!&=&\! g_{\mu\nu}
\tilde T^{(2)\alpha\beta}_{[\gamma\chi_{c2}] }
 P^{(2)}_{\alpha\beta\lambda\sigma} (p_{\chi_{c2}})
 \tilde t^{(2)\lambda\sigma}_{(12)}
f_{(12)}^{(f_2)} f_{(34)}^{(f_0)},\\
<f_0 f_2|2>\!&=&\!
 P^{(2)}_{\mu\nu\lambda\sigma} (p_{\chi_{c2}})
\tilde t^{(2)\lambda\sigma}_{(12)}
f_{(12)}^{(f_2)} f_{(34)}^{(f_0)},\\
<f_0 f_2|3>\!&=&\!
\tilde T^{(2)\alpha}_{[\gamma\chi_{c2}]\mu }
 P^{(2)}_{\nu\alpha\lambda\sigma} (p_{\chi_{c2}})
 \tilde t^{(2)\lambda\sigma}_{(12)}
f_{(12)}^{(f_2)} f_{(34)}^{(f_0)},\\
<f_2 f'_2|1>\!&=&\! g_{\mu\nu} \tilde
T^{(2)\alpha\beta}_{[\gamma\chi_{c2}] }
 P^{(2)}_{\alpha\beta\lambda\sigma} (p_{\chi_{c2}})
 \tilde t^{(2)\sigma\rho}_{(12)}
 \tilde t^{(2)\lambda}_{(34)\rho}
f_{(12)}^{(f_2)} f_{(34)}^{(f'_2)},\\
<f_2 f'_2|2>\!&=&\!
 P^{(2)}_{\mu\nu\lambda\sigma} (p_{\chi_{c2}})
\tilde t^{(2)\sigma\rho}_{(12)}
 \tilde t^{(2)\lambda}_{(34)\rho}
f_{(12)}^{(f_2)} f_{(34)}^{(f'_2)},\\
<f_2 f'_2|3>\!&=&\! \tilde T^{(2)\alpha}_{[\gamma\chi_{c2}]\mu }
 P^{(2)}_{\nu\alpha\lambda\sigma} (p_{\chi_{c2}})
\tilde t^{(2)\sigma\rho}_{(12)}
 \tilde t^{(2)\lambda}_{(34)\rho}
 \tilde t^{(2)}_{(34)\eta\tau}
f_{(12)}^{(f_2)} f_{(34)}^{(f'_2)}.
\end{eqnarray}
\subsection{$\psi\to\gamma\chi_{c0}\to\gamma \pi^+\pi^-\pi^+\pi^-$}
We construct the amplitudes $U_{\mu\nu}^i$ with a notation similar
to the  $\psi\to\gamma\chi_{c0}\to\gamma K^+ K^-\pi^+\pi^-$
channel. Here we denote $\pi^+$, $\pi^-$, $\pi^+$, $\pi^-$ as 1,
2, 3, 4.
\begin{eqnarray}
<f_0f_0|1> \!&=&\! g_{\mu\nu}[f_{(12)}^{(f_0)}f_{(34)}^{(f_0)} +
                        \{2\leftrightarrow 4\} ], \\
<f_0f_2|1> \!&=&\! g_{\mu\nu} [\tilde
T_{[f_0^{(12)}f_2^{(34)}]}^{(2)\alpha\beta} \tilde
t_{(34)\alpha\beta}^{(2)}f_{(12)}^{(f_0)}f_{(34)}^{(f_2)}
\nonumber\\ & & +\{1\leftrightarrow 3\} +\{2\leftrightarrow 4\}
+\{1\leftrightarrow 3 ~\&~
2\leftrightarrow 4\}], \\
<f_2f_2|1> \!&=&\! g_{\mu\nu} [f_{(12)}^{(f_2)}f_{(34)}^{(f_2)}
\tilde t_{(12)}^{(2)\alpha\beta}\tilde t_{(34)\alpha\beta}^{(2)}
+\{2\leftrightarrow 4\}], \\
<f_2f_2|2> \!&=&\! g_{\mu\nu} [f_{(12)}^{(f_2)}f_{(34)}^{(f_2)}
\tilde T_{[f_2^{(12)}f_2^{(34)}]}^{(2)\alpha\beta} \tilde
t_{(12)\alpha}^{(2)\gamma} \tilde t_{(34)\beta\gamma}^{(2)}
+\{2\leftrightarrow 4\}], \\
<\rho\rho|1>  \!&=&\! g_{\mu\nu}
[f_{(12)}^{(\rho)}f_{(34)}^{(\rho)} \tilde
t_{(12)}^{(1)\alpha}\tilde t_{(34)\alpha}^{(1)}
+\{2\leftrightarrow 4\}], \\
<\rho\rho|2>  \!&=&\! g_{\mu\nu}
[f_{(12)}^{(\rho)}f_{(34)}^{(\rho)} \tilde
T_{[\rho(12)\rho(34)]}^{(2)\alpha\beta} \tilde
t_{(12)\alpha}^{(1)}\tilde t_{(34)\beta}^{(1)}
+\{2\leftrightarrow 4\}], \\
 <\pi\pi^{\prime}|\pi\sigma> \!&=&\! g_{\mu\nu}
                         [f_{(123)}^{(\pi^{\prime})}
                         (f_{(12)}^{(\sigma)} + f_{(32)}^{(\sigma)}) +
                          f_{(234)}^{(\pi^{\prime})}
                         (f_{(23)}^{(\sigma)} + f_{(34)}^{(\sigma)}) +
                         \nonumber \\
                   &   & f_{(143)}^{(\pi^{\prime})}
                         (f_{(14)}^{(\sigma)} + f_{(34)}^{(\sigma)}) +
                          f_{(214)}^{(\pi^{\prime})}
                         (f_{(21)}^{(\sigma)} +
                         f_{(14)}^{(\sigma)})],
                         \\
<\pi\pi^{\prime}|\pi\rho>
\!&=&\!g_{\mu\nu}[f_{(123)}^{(\pi')}f_{(12)}^{(\rho)} \tilde
t^{(1)\alpha}_{(\rho 3)} \tilde t^{(1)}_{(12)\alpha} +
f_{(234)}^{(\pi')}f_{(23)}^{(\rho)} \tilde t^{(1)\alpha}_{(\rho
4)} \tilde t^{(1)}_{(23)\alpha} + \nonumber \\ & &
+\{1\leftrightarrow 3\}+\{2\leftrightarrow 4\}
+\{1\leftrightarrow 3 ~\&~ 2\leftrightarrow 4\}], \\
<\pi a_{1}|\pi\sigma> \!&=&\!g_{\mu\nu}
[f_{(123)}^{(a_1)}f_{(12)}^{(\sigma)} \tilde
T^{(1)\alpha}_{(a_14)} \tilde t^{(1)}_{(\sigma 3)\alpha} +
f_{(234)}^{(a_1)}f_{(23)}^{(\sigma)} \tilde T^{(1)\alpha}_{(a_11)}
\tilde t^{(1)}_{(\sigma 4)\alpha} +
                         \nonumber \\ & &
+\{1\leftrightarrow 3\}+\{2\leftrightarrow 4\}
+\{1\leftrightarrow 3 ~\&~ 2\leftrightarrow 4\}], \\
<\pi a_{1}|\pi\rho> \!&=&\!g_{\mu\nu}
[P^{(1)}_{\alpha\beta}(p_{(123)})\tilde T^{(1)\alpha}_{(a_14)}
\tilde t^{(1)\beta}_{(12)} f_{(123)}^{(a_1)}f_{(12)}^{(\rho)}
\nonumber\\& & +P^{(1)}_{\alpha\beta}(p_{(234)}) \tilde
T^{(1)\alpha}_{(a_11)} \tilde t^{(1)\beta}_{(23)}
f_{(234)}^{(a_1)}f_{(23)}^{(\rho)}
                         \nonumber \\ & &
+\{1\leftrightarrow 3\}+\{2\leftrightarrow 4\} +\{1\leftrightarrow
3 ~\&~ 2\leftrightarrow 4\}].
\end{eqnarray}

\subsection{$\psi\to\gamma\chi_{c1}\to\gamma \pi^+\pi^-\pi^+\pi^-$ }
In this subsection we construct the amplitudes $U_{\mu\nu}^i$ for
the process $\psi\to\gamma\chi_{c1}\to\gamma
\pi^+\pi^-\pi^+\pi^-$.  The most possible intermediate states are:
  $f_0 f_2$,
$f_2 f_2$, and $\rho\rho$ with $f_0 $, $f_2$, and $\rho \to
\pi^+\pi^-$.

\begin{eqnarray}
<f_0 f_2|1> \!&=&\! \varepsilon_{\mu\nu\alpha\beta} p^\beta_{\psi}
\varepsilon_{\lambda\gamma\delta\xi} p^\xi_{\chi_{c1}}
g^{\alpha\delta}
 [
 \tilde T^{(2)\gamma}_{[f_0 \bar f_2] \sigma} \tilde t_{(12)}^{(2)\lambda\sigma}
f_{(34)}^{(f_0)}f_{(12)}^{(f_2)} \nonumber\\
 && +\{1\leftrightarrow 3\}+\{2\leftrightarrow 4\} +\{1\leftrightarrow
3 ~\&~ 2\leftrightarrow 4\}] ,\\
<f_0 f_2|2> \!&=&\! \varepsilon_{\rho\nu\alpha\beta}
p^\beta_{\psi} q^\rho q_\mu \varepsilon_{\lambda\gamma\delta\xi}
p^\xi_{\chi_{c1}}  g^{\alpha\delta} [
 \tilde T^{(2)\gamma}_{[f_0 \bar f_2] \sigma} \tilde t_{(12)}^{(2)\lambda\sigma}\nonumber\\
 &&
f_{(34)}^{(f_0)}f_{(12)}^{(f_2)} +\{1\leftrightarrow
3\}+\{2\leftrightarrow 4\} +\{1\leftrightarrow
3 ~\&~ 2\leftrightarrow 4\}] ,\\
<f_2 f_2|1> \!&=&\! \varepsilon_{\mu\nu\alpha\beta} p^\beta_{\psi}
\varepsilon_{\lambda\gamma\delta\xi} p^\xi_{\chi_{c1}}
g^{\alpha\delta} [
 \tilde t_{(12)}^{(2)\lambda\sigma} \tilde t^{(2)\gamma}_{(34)\sigma}
f_{(12)}^{(f_2)}f_{(34)}^{(f_2)}  +\{2\leftrightarrow 4\}] ,\\
<f_2 f_2|2> \!&=&\! \varepsilon_{\rho\nu\alpha\beta}
p^\beta_{\psi} q^\rho q_\mu \varepsilon_{\lambda\gamma\delta\xi}
p^\xi_{\chi_{c1}} g^{\alpha\delta}[
 \tilde t_{(12)}^{(2)\lambda\sigma} \tilde t^{(2)\gamma}_{(34)\sigma}
f_{(12)}^{(K^*_2)}f_{(34)}^{(f_2)} +\{2\leftrightarrow 4\}],\\
<\rho \rho|1> \!&=&\! \varepsilon_{\mu\nu\alpha\beta}
p^\beta_{\psi} \varepsilon_{\lambda\sigma\gamma\delta}
p^\delta_{\chi_{c1}}
   g^{\alpha\gamma}
[ \tilde t^{(1)\lambda}_{(12)} \tilde t^{(1)\sigma}_{(34)}
f_{(12)}^{(\rho)}f_{(34)}^{(\rho)} +\{2\leftrightarrow 4\}],\\
<\rho \rho|2> \!&=&\! \varepsilon_{\xi\nu\alpha\beta}
p^\beta_{\psi} q^\xi q_\mu \varepsilon_{\lambda\sigma\gamma\delta}
p^\delta_{\chi_{c1}}
   g^{\alpha\gamma}
 [\tilde t^{(1)\lambda}_{(12)} \tilde t^{(1)\sigma}_{(34)}
f_{(12)}^{(\rho)}f_{(34)}^{(\rho)} +\{2\leftrightarrow 4\}].
\end{eqnarray}

\subsection{$\psi\to\gamma\chi_{c2}\to\gamma \pi^+\pi^-\pi^+\pi^-$ }
 The
most possible intermediate states are $f_0 f_0$, $f_0 f_2$, $f_2
f_2$, and $\rho\rho$ with $f_0 $, $f_2$, $\rho\to \pi^+\pi^-$.
Then we have the following convariant tensor amplitudes :
\begin{eqnarray}
<f_0f_0|1> \!&=&\! g_{\mu\nu}\tilde
T^{(2)\alpha\beta}_{[\gamma\chi_{c2}]}  \tilde T^{(2)}_{ [f_0 f_0
]\alpha\beta }
 [f_{(12)}^{(f_0)}f_{(34)}^{(f_0)}
                      +  \{2\leftrightarrow 4\} ]
                       , \\
<f_0f_0|2> \!&=&\! \tilde T^{(2)}_{ [f_0 f_0 ]\mu\nu}
 [f_{(12)}^{(f_0)}f_{(34)}^{(f_0)} +
                        \{2\leftrightarrow 4\} ]
                       , \\
<f_0f_0|3> \!&=&\!  \tilde T^{(2)\alpha}_{[\gamma \chi_{c2}]\mu}
\tilde T^{(2)}_{[ f_0 f_0 ]\nu \alpha}
 [f_{(12)}^{(f_0)}f_{(34)}^{(f_0)} +
                        \{2\leftrightarrow 4\} ]
                      ,\\
<f_0 f_2|1> \!&=&\! g_{\mu\nu} \tilde
T^{(2)\alpha\beta}_{[\gamma\chi_{c2}]}
 P^{(2)}_{\alpha\beta\lambda\sigma}(p_{\chi_{c2}})
 [
 \tilde t_{(12)}^{(2)\lambda\sigma}
f_{(12)}^{(f_2)}f_{(34)}^{(f_0)}  \nonumber \\ & &
+\{1\leftrightarrow 3\}+\{2\leftrightarrow 4\} +\{1\leftrightarrow
3 ~\&~ 2\leftrightarrow 4\}]  , \\
<f_0 f_2|2> \!&=&\!
 P^{(2)}_{\mu\nu\lambda\sigma}(p_{\chi_{c2}})
 [
 \tilde t_{(12)}^{(2)\lambda\sigma}
f_{(12)}^{(f_2)}f_{(34)}^{(f_0)}  \nonumber \\ & &
+\{1\leftrightarrow 3\}+\{2\leftrightarrow 4\} +\{1\leftrightarrow
3 ~\&~ 2\leftrightarrow 4\}]  , \\
<f_0 f_2|3> \!&=&\! \tilde T^{(2)\alpha}_{[\gamma\chi_{c2}]\mu}
 P^{(2)}_{\nu\alpha\lambda\sigma}(p_{\chi_{c2}})
 [
 \tilde t_{(12)}^{(2)\lambda\sigma}
f_{(12)}^{(f_2)}f_{(34)}^{(f_0)}  \nonumber \\ & &
+\{1\leftrightarrow 3\}+\{2\leftrightarrow 4\} +\{1\leftrightarrow
3 ~\&~ 2\leftrightarrow 4\}],\\
<f_2f_2|1> \!&=&\! g_{\mu\nu} \tilde
T^{(2)\alpha\beta}_{[\gamma\chi_{c2}]}
  P^{(2)}_{\alpha\beta\lambda\sigma} (p_{\chi_{c2}})
[ \tilde t^{(2)\sigma\rho}_{(12)}
 \tilde t_{(34)\rho}^{(2)\lambda}f_{(12)}^{(f_2)}f_{(34)}^{(f_2)}
 +
                        \{2\leftrightarrow 4\} ] ,\\
<f_2f_2|2> \!&=&\!
  P^{(2)}_{\mu\nu\lambda\sigma} (p_{\chi_{c2}})
[ \tilde t^{(2)\sigma\rho}_{(12)}
 \tilde t_{(34)\rho}^{(2)\lambda}f_{(12)}^{(f_2)}f_{(34)}^{(f_2)}
 +
                        \{2\leftrightarrow 4\} ] ,\\
<f_2f_2|3> \!&=&\! \tilde T^{(2)\alpha}_{[\gamma\chi_{c2}]\mu}
  P^{(2)}_{\nu\alpha\lambda\sigma} (p_{\chi_{c2}})
[ \tilde t^{(2)\sigma\rho}_{(12)}
 \tilde t_{(34)\rho}^{(2)\lambda}f_{(12)}^{(f_2)}f_{(34)}^{(f_2)}
 +
                        \{2\leftrightarrow 4\} ] ,\\
<\rho\rho|1> \!&=&\! g_{\mu\nu} \tilde
T^{(2)\alpha\beta}_{[\gamma\chi_{c2}]}
 P^{(2)}_{\alpha\beta\lambda\sigma}(p_{\chi_{c2}})
[\tilde t^{(1)\lambda}_{(12)}
 \tilde t_{(34)}^{(1)\sigma}
f_{(12)}^{(\rho)}f_{(34)}^{(\rho)}   +
                        \{2\leftrightarrow 4\} ],\\
<\rho\rho|2> \!&=&\!
 P^{(2)}_{\mu\nu\lambda\sigma}(p_{\chi_{c2}})
[\tilde t^{(1)\lambda}_{(12)}
 \tilde t_{(34)}^{(1)\sigma}
f_{(12)}^{(\rho)}f_{(34)}^{(\rho)}   +
                        \{2\leftrightarrow 4\} ],\\
<\rho\rho|3> \!&=&\!  \tilde T^{(2)\alpha}_{[\gamma\chi_{c2}]\mu}
 P^{(2)}_{\nu\alpha\lambda\sigma}(p_{\chi_{c2}})
[\tilde t^{(1)\lambda}_{(12)}
 \tilde t_{(34)}^{(1)\sigma}
f_{(12)}^{(\rho)}f_{(34)}^{(\rho)}   +
                        \{2\leftrightarrow 4\} ].
\end{eqnarray}
Here $f_0$, $f_2$ and $\rho$ can be replaced by any $f'_0$, $f'_2$
and $\rho'$, respectively.

\section{Conclusion}
First of all, we provide a theoretical PWA formalism for the
radiative decay $J/\psi \to \gamma p \bar p$, which is also
applicable to the processes $J/\psi \to \gamma\Lambda\bar \Lambda,
\gamma\Sigma \bar\Sigma$ and $\gamma\Xi\bar \Xi$.  Then we present
a general covariant formalism for the PWA of the double radiative
decay  $\psi \to \gamma\gamma V(\rho, \omega, \phi)$ processes.
Finally, we give the PWA formulae for $\psi(2s)$ radiative decays
into $K^+K^-\pi^+\pi^-$ and $\pi^+\pi^-\pi^+\pi^-$  that are very
useful to study $\chi_{cJ}$ charmonium states. We have constructed
most possible covariant tensor amplitudes for intermediate
resonant states of $J\leq 2$. For intermediate resonant states of
$J\geq 3$, the production vertices need $L\geq 2$  and are
expected to be suppressed \cite{zou}. The formulae here can be
directly used to perform partial wave analysis of forthcoming high
statistics data from CLEO-c and BES-III on these channels to
extract useful information on the baryon-antibaryon interactions,
and  $\psi \to \gamma\gamma V(\rho, \omega, \phi)$
 processes to extract information on the
flavor content of any meson resonances (R) with positive charge
parity ($C=+$) and mass above 1 GeV, as well as
$\psi(2s)\to\gamma\chi_{cJ}$ with $\chi_{cJ}$ decays into
$K^+K^-\pi^+\pi^-$ and $\pi^+\pi^-\pi^+\pi^-$ to study gluon
hadronization dynamics.

\section{Acknowledgements}
We thank Z.J.Guo and C.Z.Yuan for useful discussions. The work of
S.Dulat is supported by the National Natural Sciences Foundation
of China under Grant No.10465004, and partly
 by the Abdus Salam
International Centre for Theoretical Physics, Trieste, Italy, as
well as by Xinjiang University. The work of B.S. Zou is partly
supported by CAS Knowledge Innovation Project (KJCX2-SW-N02) and
the National Natural Sciences Foundation of China under Grant Nos.
10225525,10435080.

\end{document}